\documentclass[12pt]{article}
\pdfoutput=1
\usepackage{graphicx}
\usepackage{color,amssymb,amsmath}
\usepackage[a4paper, total={16.5cm,22cm}]{geometry}
\usepackage{cite}
\usepackage{verbatim}
\usepackage{fancyvrb}
\usepackage[normalem]{ulem}

\DefineVerbatimEnvironment{Code}{BVerbatim}{baseline=t}
\usepackage[colorlinks=true,linkcolor=dark_blue,
  citecolor=dark_red,urlcolor=dark_red,linktocpage=false]{hyperref}
\definecolor{dark_red}{rgb}{0.6,0,0}
\definecolor{dark_blue}{rgb}{0,0,0.6}

\usepackage{slashed}

\begin{document}

\newcommand{\lrf}[2]{ \left(\frac{#1}{#2}\right)}
\newcommand{\lrfp}[3]{ \left(\frac{#1}{#2} \right)^{#3}}
\newcommand{\vev}[1]{\langle #1\rangle}
\newcommand{\hf}{\mathchar`-}

\def\beq{\begin{equation}}
\def\eeq{\end{equation}}
\def\beqn{\begin{eqnarray}}
\def\eeqn{\end{eqnarray}}

\def\gluino{\widetilde g}
\def\grav{\widetilde G}
\def\tone{\widetilde t_1}
\def\none{\widetilde \chi_1^0}
\def\nonetwo{\widetilde \chi_{1/2}^0}
\def\cpone{\widetilde \chi_1^+}
\def\cmone{\widetilde \chi_1^-}
\def\cone{\widetilde \chi_1^\pm}

\begin{titlepage}

\begin{center}

\vskip 2cm

{\Large \bf 
Gravitino vs Neutralino LSP at the LHC
}

\vskip 1.5cm

{
Jong Soo Kim$^{(a)}$,
Stefan Pokorski$^{(b)}$, 
Krzysztof Rolbiecki$^{(b)}$,
Kazuki Sakurai$^{(b)}$
}

\vskip 0.7cm

$^{(b)}${\em National Institute for Theoretical Physics and \\
University of the Witwatersrand, Johannesburg, Wits 2050, South Africa
}

\vskip 0.3cm

$^{(b)}${\em
Institute of Theoretical Physics, Faculty of Physics,\\University of Warsaw, ul.~Pasteura 5, PL-02-093 Warsaw, Poland\\[0.1cm]
}

\end{center}

\vskip 1.5cm
\begin{abstract}
Using the latest LHC data, we analyse and compare the lower limits on the masses of gluinos and the lightest stop in two natural supersymmetric motivated scenarios:  one with a neutralino  being the lightest supersymmetric particle (LSP) and the other one with gravitino as the LSP and neutralino as the next-to-lightest supersymmetric particle. In the second case our analysis applies to neutralinos promptly decaying to very  light  gravitinos, which are of cosmological interest, and are generic for low, of order $\mathcal{O}(100)$~TeV,  messenger scale in gauge mediation models. We find that the lower bounds on the  gluino and the lightest stop  masses are stronger for the gravitino LSP scenarios
due to the extra handle from the decay products of neutralinos. Generally, in contrast to the neutralino LSP case the limits now extend to a region of compressed spectrum. In bino scenarios the highest excluded stop mass increases from 1000~GeV to almost 1400~GeV.  
Additionally, in the higgsino-like NLSP scenario the higgsinos below 650~GeV are universally excluded and the stop mass limit is $m_{\tilde{t}} > 1150$~GeV, whereas there is no limit on stops in the higgsino LSP model for $m_{\tilde{h}} = 650$~GeV. Nevertheless, we find that the low messenger scale still ameliorates the fine tuning in the electroweak potential. 
\end{abstract}

\end{titlepage}

\renewcommand{\thepage}{\arabic{page}}
\setcounter{page}{1}
\renewcommand{\thefootnote}{$\natural$\arabic{footnote}}
\setcounter{footnote}{0}

\newpage

\section{Introduction} 

In this paper we analyse the latest LHC data, to obtain and compare the  bounds on the gluino and the lightest stop masses in two general scenarios, one with neutralino as the lightest supersymmetric particle (LSP) and the other one with gravitino as the LSP. Focusing on the  bounds for gluino and the lightest  stop is motivated by  the question of naturalness of the Higgs potential in supersymmetric (SUSY) models. It has been well appreciated already for many decades that those particle masses, together  with the higgsino mass, are most crucial for the degree of fine tuning; see Ref.~\cite{Papucci:2011wy, Buckley:2016kvr} for a recent analysis. Our analysis is performed in the framework  of three simplified models described in detail in Section~\ref{sec:model}. The main assumption here is that only the stop, gluino and the neutralino LSP (and the gravitino in the second scenario) are relevant for the production  and decay processes leading to the collider signatures under study. It is assumed that the other superpartners decouple because they are either heavy  or have small cross sections. It should be stress that such a situation indeed occurs in many explicit models studied in the literature. The three simplified models correspond to the lightest neutralino being pure bino, wino or higgsino. In the scenario with neutralino LSP, the bounds on the the gluino and stop masses have been extensively studied using the LHC data in experimental and theoretical papers, see e.g.\ Refs.~\cite{1709_04183,1712_02332,ATLAS-CONF-2018-041,Sirunyan:2018ell,Sirunyan:2018vjp,CMS-PAS-SUS-19-005,Baer:2017yqq}. We  revisit this class of simplified models taking into account recent LHC results and directly compare with the gravitino LSP scenarios.

For the case with gravitino LSP, we assume that neutralino is the next-to-lightest supersymmetric particle (NLSP) particle decaying promptly, with $c \tau_{\tilde \chi_1^0} \lesssim 1$\,mm, and focus  on the corresponding LHC signatures\footnote{For a recent analysis with a long-lived NLSP see e.g.\ Ref.~\cite{Knapen:2016exe}.} \cite{Feng:2010ij,Kim:2017pvm} (for a similar analysis  based on the Tevatron data see Ref.~\cite{Meade:2009qv, Ruderman:2011vv}). It follows from this assumption that the obtained bounds apply when gravitinos are very light, with masses $m_{3/2} \lesssim 1$\,keV for neutralinos lighter than ${\cal O}(1)$TeV\ (see Section~\ref{sec:model} for details). For substantially heavier neutralinos it is not possible to apply our recasting strategy since the neutralino becomes long-lived. 

Gravitino LSP is generic for gauge mediated SUSY  breaking (GMSB) scenarios. Its mass reads $m_{3/2} = \frac{F}{\sqrt{3} M_{\rm Pl}}$, where $F$ is the supersymmetry breaking $F$-term (or their combination), so the gravitino mass is model dependent. In general, gaugino masses in the GMSB scenario are given by one-loop contributions, roughly   of the order $m_{\rm SUSY} \sim 0.01 \frac{F}{\Lambda}$. This means the messenger scale $\Lambda$ is linearly related to $m_{3/2}$, for a fixed $m_{\rm SUSY}$, and the mass range of gravitinos considered in this paper is consistent with the region $\Lambda \lesssim $\,100--1000\,TeV. Very light gravitinos are theoretically interesting for several reasons. One motivation is to relax the apparent fine-tuning in the Higgs sector thanks to  low values of the messenger scale.\footnote{Low cut-off scale in the loops contributing to the Higgs potential is also possible in the neutralino LSP scenarios and heavy gravitinos, e.g.\ in the SUSY twin Higgs models. This case is not analysed in this paper, see however~\cite{Craig:2013fga}.}

Very light gravitino, $m_{3/2} \lesssim {\cal O}(10)$\,eV, is also motivated by cosmology since it can evade the gravitino problem \cite{Moroi:1993mb,Rychkov:2007uq} and simultaneously allow for very high reheating temperature, convenient for leptogenesis. Lighter gravitino implies stronger interaction between the spin-$\frac{1}{2}$ Goldstino component and the MSSM particles, and such gravitinos are easily thermalised if the reheating temperature is high enough. Once the gravitino reaches thermal equilibrium there will be no upper limit on the reheating temperature since the abundance is determined by the freeze-out temperature rather than the reheating temperature. In this case, the relic abundance of gravitinos can be approximately estimated in terms of the relativistic degrees of freedom, $g_{*3/2} \sim 100$, at the gravitino decoupling epoch as (see e.g.~\cite{Osato:2016ixc})
\begin{equation}
\Omega_{3/2} h^2 ~\sim~ 0.1 \, \Big( \frac{m_{3/2}}{100\,{\rm eV}} \Big) \Big( \frac{90}{g_{*3/2}} \Big) \,.
\label{eq:thermal}
\end{equation}
Thus, the relic abundance of gravitinos, that were originally in thermal equilibrium,  exceeds the observed  abundance of dark matter (DM) for $m_{3/2} \gtrsim 100$\,eV. On the other hand, fits to the matter fluctuations at small scales (small scale structure formation) put a lower bound on  the mass of a warm dark matter particle, $m_{3/2} \gtrsim {\cal O}(1)$\,keV, if it fully accounts for the observed DM abundance~\cite{Viel:2005qj}. Therefore,  the light thermal gravitinos cannot be the dominant component of the dark matter. Assuming that DM consists of the light thermal gravitinos and some additional cold dark matter (CDM) constituent, $\Omega_{\rm DM}=\Omega_{\rm CDM}+\Omega_{3/2}$, the constraint from the Ly-$\alpha$ and  sets the upper bound, $m_{3/2} < 16$\,eV~\cite{Viel:2005qj}. A more recent study~\cite{Osato:2016ixc} sets the tighter upper bound, $m_{3/2} < 4.7$\,eV, using the data of the cosmic microwave background (CMB) lensing collected by Planck and of cosmic shear measured by the Canada-France-Hawaii Lensing Survey, combined with analyses of the primary CMB anisotropies and the baryon acoustic oscillations in galaxy distributions. These studies assume that the whole relic gravitinos come from the thermal freeze-out. However, gravitinos could be produced additionally, for example, from decays of the NLSP, after gravitino freeze-out, which increases gravitino abundance as $\Omega_{3/2}^{\rm decay} = \frac{m_{3/2}}{m_{\rm NLSP}} \Omega_{\rm NLSP}$.\footnote{This follows from the fact that a single grativino is produced per NLSP decay, preserving the number density.} For light gravitinos this contribution is negligible in the realistic parameter region. However, if there is additional gravitino production mechanism that is significant, the above bounds become even tighter.

The light thermal gravitino cannot be the dominant component of the dark matter  but remains cosmologically interesting for the reasons mentioned above.\footnote{In such a case, we must invoke another particle as a main component of the dark matter. The invisible axion may be one of such candidates.} We emphasise that there is a room for that argument to be modified. For example, if the reheating temperature is very low ($T_R \ll {\cal O}(100)$ GeV), gravitinos will not reach the thermal equilibrium, and the relic abundance can be much smaller than what is obtained from Eq.~\eqref{eq:thermal} \cite{Moroi:1993mb}. In this case, the above limits \cite{Viel:2005qj, Osato:2016ixc} will not apply. The whole mass range considered in this paper is, therefore, in principle allowed by cosmological considerations.

Throughout this paper we assume for simplicity the conservation of $R$-parity so that the gravitino is absolutely stable.  However, this assumption can be relaxed as long as decay branching ratios of sparticles in our analysis are not modified. It has been discussed that even if $R$-parity is broken the lifetime of gravitinos can be longer than the age of the Universe and the gravitino can be a viable dark matter candidate \cite{Takayama:2000uz}. For a relatively heavy gravitino ($m_{3/2} \sim 100$ GeV), it has been argued the decay products of relic gravitino may be found in cosmic rays (see e.g.\ Refs.~\cite{Takayama:2000uz, Bertone:2007aw, Ibarra:2007wg, Covi:2008jy, Choi:2010jt}). However, it is not realistic to expect such a signature for the light gravitino ($m_{3/2} \lesssim 1$ keV) in our analysis.

\section{The light gravitino and model setup\label{sec:model}} 

Our study is organized around the question of how the presence of gravitino as the lightest supersymmetric particle ---to this end at least from a collider point of view--- changes exclusion limits at the LHC. We focus on models that have been motivated by the concept of natural SUSY, hence we consider spectra where among light particles we have gauginos/higgsinos, serving as the (next-to-lightest) supersymmetric particle, stops and gluinos. Other SUSY particles are assumed to be heavy, i.e.\ outside the collider reach and with a negligible impact on decay chains of light particles.  

We examine three types of the simplified models:
\begin{itemize}
 \item a stop model, where gluinos are decoupled;
 \item a gluino model, where stops are assumed to be heavy, however the gluino decays are mediated by off-shell stops;
 \item a stop-gluino model where both particles are within or close to kinematic reach of the LHC and no definite mass hierarchy is assumed.
\end{itemize}
In all the models the strongly interacting particles are accompanied by electroweakinos: either the higgsino doublets, wino triplet or bino singlet -- which are assumed to be lighter than the stops and gluinos. Finally, the models can be completed by adding the light gravitino. Generally the mass hierarchy is as follows:
\begin{equation}
  m_{\gluino}, m_{\tone} > m_{\widetilde{\chi}}\ ( > m_{3/2} ).
\end{equation}
These models are parametrised by the masses of participating particles: $m_{\gluino}$, $m_{\tone}$, $m_{\widetilde \chi}$. Since we assume that only one type of electroweakinos is present at a time, the mixing between gaugino-higgsino states is neglected, which implies that the additional charginos and neutralinos are almost mass degenerate with the lightest neutralino. 

\subsection{The electroweakino sector}
The electroweakino sector is characterised by soft SUSY-breaking masses of bino and wino, $M_1$ and $M_2$, respectively, the higgsino mass $\mu$ and the ratio of Higgs vacuum expectation values $\tan\beta$. The neutralino gauge eigenstates $\widetilde  \psi^0 = (\widetilde B, \widetilde W^0, \widetilde H_d, \widetilde H_u)^T$, are related to mass eigenstates by means of the neutralino mixing matrix $N_{ij}$; see e.g.\ Ref.~\cite{Haber:1984rc,Choi:2001ww} for details. 

For our purposes the most important property of the lightest neutralino is its gaugino-higgsino composition. While this can be completely general, in our simplified model setup we only have one type of light electroweakinos. In such a case, the mixing will be small and can be neglected. Therefore, we classify our models according to the dominant component of the NLSP in three types, defined in terms of the neutralino mixing components as\footnote{Strictly speaking, the higgsino mixing depends on the sign of $\mu$ and EW gaugino masses as $N_{13} = - \eta N_{14} = 1/\sqrt{2}$ with $\eta \equiv {\rm sign}(\mu) \cdot {\rm sign}\left( \frac{M_1}{M_2} + \tan^2 \theta_W \right)$. However, the effect of this is small for a moderate or large $\tan\beta$ regime. In our study we take $\eta = 1$ for simplicity.}
\begin{eqnarray}
{\rm bino}&:& N_{11} \approx 1,~ N_{1i} \approx 0~ (i \neq 1) 
\nonumber \\
{\rm wino}&:& N_{12} \approx 1,~ N_{1i} \approx 0~ (i \neq 2) 
\nonumber \\
{\rm higgsino}&:& N_{13} \approx - N_{14} \approx 1/\sqrt{2},~ N_{1i} \approx 0~ (i = 1,2) .
\label{eq:neu1component}
\end{eqnarray}
For each of these cases a different decay pattern emerges which additionally depends on the NLSP mass and, for higgsinos, on $\tan\beta$. 

Taking the bino mass $M_1$ much smaller than the wino, $M_2$, and higgsino, $\mu$, masses we obtain the bino scenario. It means that the relevant electroweakino sector consist of just one lightest neutralino. Because of its vanishing coupling to the gauge bosons the bino mass is currently unconstrained in collider experiments.    

The wino scenario is realized when $M_2 \ll M_1, |\mu|$, there are two charged states together with the neutralino, which we write $\widetilde W^+, \widetilde W^-$ and $\widetilde W^0$, respectively. Since they form an SU(2) triplet in the limit $v \to 0$, they are almost mass degenerate. The charged and neutral states will split after including the electroweak symmetry breaking effects as well as the radiative correction. The size of this mass splitting is in general small and irrelevant from the collider perspective, unless it is so small that the charged state becomes long-lived. In such a situation, the long-lived charged wino gives rise to so-called disappearing track signature. The ATLAS and CMS collaborations have analysed such a signature and put a strong constraint on the long-lived wino, $m_{\widetilde W^+} \gtrsim 460$\,GeV \cite{Aaboud:2017mpt,Sirunyan:2018ldc}. In this paper, we do not consider such a case and assume that the charged winos are short-lived at the LHC.

For $|\mu| \ll M_1, M_2$ we obtain the higgsino scenario for which there are two neutral and two charged states coming from four components of the two higgsino doublets, $\widetilde h^0_1, \widetilde h^0_2, \widetilde h^+, \widetilde h^-$.  Up to the electroweak symmetry breaking and radiative corrections, they are nearly mass degenerate. On the other hand, the mass splitting among the higgsino states is in general larger than that for winos, and we again assume that the higgsinos are short-lived in our analysis.

\subsection{NLSP decay to gravitino}

\begin{figure}[t]
    \begin{minipage}{0.5\hsize}
        \begin{center}
        \includegraphics[width=8cm]{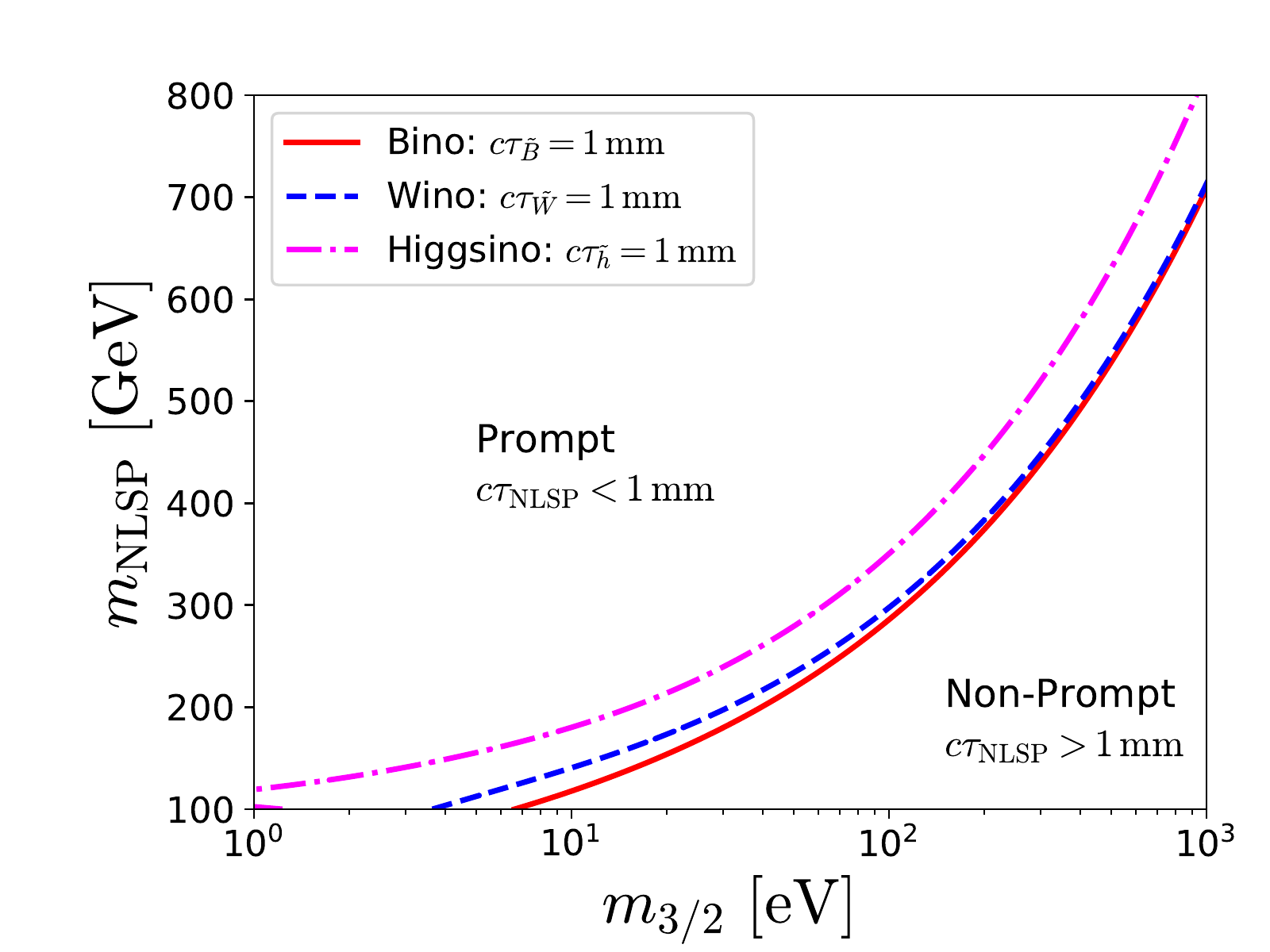}
        \end{center}
    \end{minipage}
    \begin{minipage}{0.5\hsize}
        \begin{center}
        \includegraphics[width=8cm]{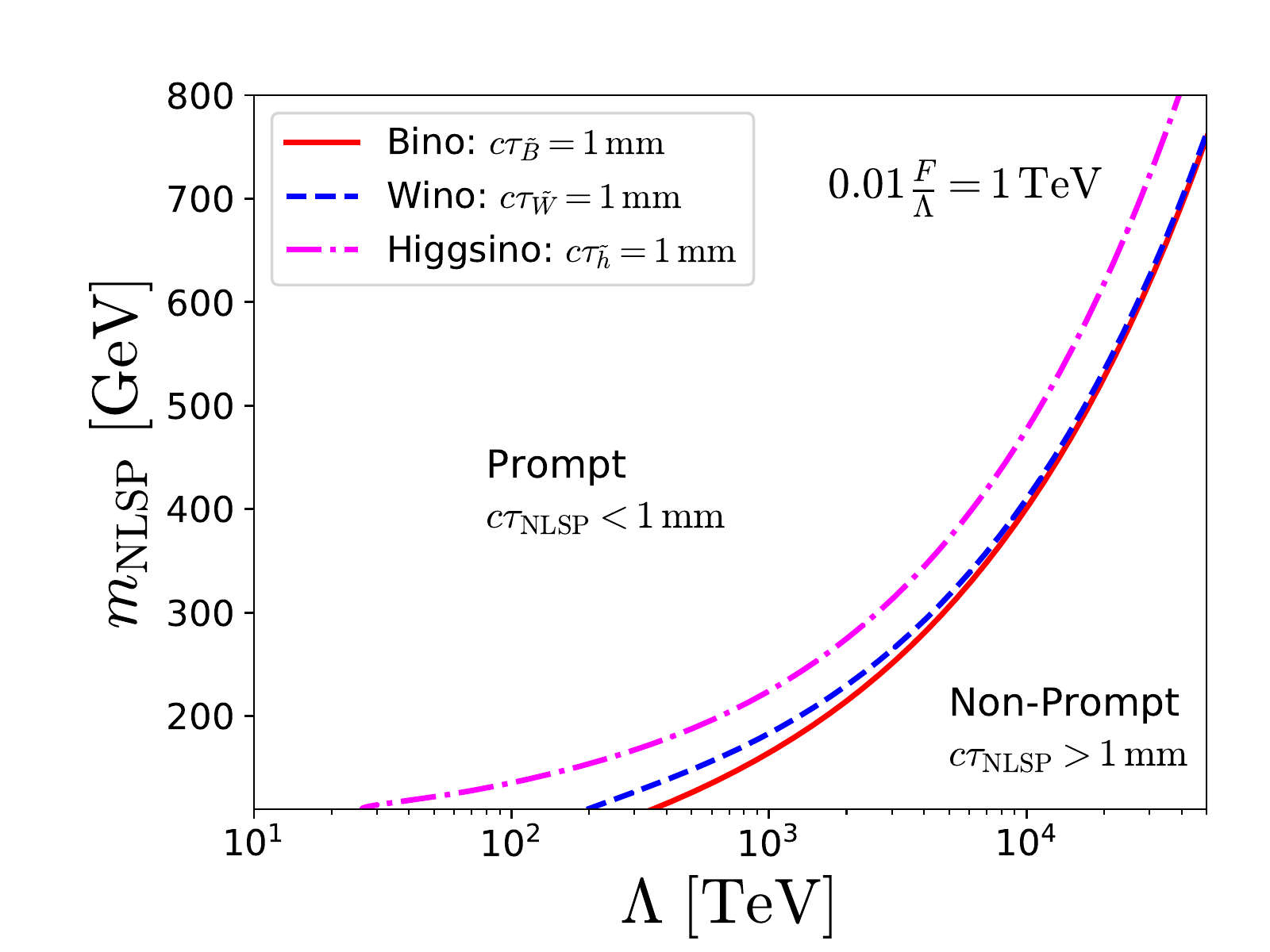}
        \end{center}
    \end{minipage}
    \caption{
    \small Left: $c \tau_{\tilde \chi_1^0} = 1$\,mm contours in the $m_{3/2}$ vs $m_{\widetilde \chi_1^0}$ plane.
    Right: $c \tau_{\tilde \chi_1^0} = 1$\,mm contours in the $\Lambda$ vs $m_{\widetilde \chi_1^0}$ plane
    assuming $0.01 \frac{F}{\Lambda} = m_{\rm SUSY} = 1$\,TeV.
    In both plots, the red-solid, blue-dashed and pink-dot-dashed curves correspond to the bino, wino and higgsino NLSPs, respectively. 
    The regions above the contours satisfy the assumption that the NLSP neutralino decays promptly into a gravitino, $c \tau_{\tilde \chi_1^0} < 1$\,mm.  In the lower right region, the NSLP is long-lived and our analysis may not be applied.  
    \label{fig:lifetime}    
    }
\end{figure}

In models with the gravitino LSP we assume it is light enough so that it can be treated as  massless at the LHC. We also assume that the decay of the NLSP neutralino to gravitino is prompt, taking a conservative limit $c \tau_{\tilde \chi_1^0} \lesssim 1$\,mm, cf.\ Ref.~\cite{Chatrchyan:2012jwg,CMS-PAS-EXO-12-035,Aad:2013oua,Aad:2014gfa,Aaboud:2017iio,Hiller:2009ii}. The partial decay rates of the lightest neutralino into the gravitino are given by~\cite{Ambrosanio:1996jn, Meade:2009qv}
\begin{eqnarray}
\Gamma(\tilde \chi_1^0 \to \tilde G  \gamma) &=& \big| N_{11} c_W + N_{12} s_W \big|^2 {\cal A} \,,
\nonumber \\
\Gamma(\tilde \chi_1^0 \to \tilde G  Z) &=& \Big( \big| N_{12} c_W - N_{11} s_W \big|^2 
+ \frac{1}{2} \big| N_{13} c_\beta - N_{14} s_\beta \big|^2 \Big)
\Big( 1 - \frac{m_Z^2}{m^2_{\tilde \chi_1^0}} \Big)^4
{\cal A} \,,
\nonumber \\
\Gamma(\tilde \chi_1^0 \to \tilde G  h) &=&  \frac{1}{2} \big| N_{13} c_\beta + N_{14} s_\beta \big|^2
\Big( 1 - \frac{m_h^2}{m^2_{\tilde \chi_1^0}} \Big)^4
{\cal A} \,,
\label{eq:br_neut}
\end{eqnarray}
where $N_{ij}$ is the neutralino mixing matrix and
\begin{equation}
{\cal A} ~=~ \frac{m^5_{\tilde \chi_1^0}}{16 \pi m^2_{3/2} M^2_{\rm pl}}
\,\sim \,
\frac{1}{0.3\,\rm mm}
\Big( \frac{m_{\tilde \chi_1^0}}{100\,\rm GeV} \Big)^5
\Big( \frac{m_{3/2}}{10\,\rm eV} \Big)^{-2} \,.
\end{equation}
In the left panel of Figure~\ref{fig:lifetime} we plot contours of a fixed neutralino lifetime $c \tau_{\none} = 1$\,mm in the gravitino-neutralino mass plane. The three contours correspond to the lightest neutralino which is predominantly bino (red-solid), wino (blue-dashed) and higgsino (pink-dotted-dashed). The prompt region ($c \tau_{\none} < 1$\,mm) is located above contours (the top-left part of the plots), allowing fairly light gravitinos with $m_{3/2} \lesssim 5$\,eV--1\,keV for neutralinos lighter than $\mathcal{O}(1)$\,TeV. It justifies our assumption that the gravitino can be treated as a massless particle  in dealing with its kinematics at colliders and we conveniently fix $m_{3/2}$ to 1\,eV throughout our analysis. The right panel of Figure~\ref{fig:lifetime} recasts the calculation in the $\Lambda$--$m_{\rm NLSP}$ plane, where $\Lambda$ is the messenger scale and assuming $0.01 \frac{F}{\Lambda} = 1$\,TeV.

The branching ratio of the bino-like neutralino can be obtained by substituting $N_{1i} = \delta_{1i}$ into Eq.~\eqref{eq:br_neut}, which leads to
\begin{eqnarray}
{\rm BR}(\widetilde B \to \gamma \grav) &=& \frac{c_W^2}{c_W^2 + s_W^2\big( 1 - \frac{m_Z^2}{m^2_{\widetilde B}} \big)^4 } \,, 
\nonumber \\
{\rm BR}(\widetilde B \to Z \grav) &=& 1 - {\rm Br}(\widetilde B \to \gamma \grav) \,, 
\nonumber \\
{\rm BR}(\widetilde B \to h \grav) &=& 0 \,.
\end{eqnarray}
Numerically, the $\none \to \gamma \grav$ mode dominates the $Z$-boson mode for any bino mass. In particular in the limit $m_Z/m_{\widetilde B} \to 0$, they approach ${\rm BR}(\widetilde B \to \gamma \grav) \to c^2_W \simeq 0.77$ and ${\rm BR}(\widetilde B \to Z \grav) \to s^2_W \simeq 0.23$. Therefore the models are mainly constrained by the analysis targeting photon final states as we will see later.

For the wino-like NLSP the branching ratio of $\widetilde W^0$ is obtained by taking $N_{1i} = \delta_{2i}$ in Eq.~\eqref{eq:br_neut}, which gives
\begin{eqnarray}
{\rm BR}(\widetilde W^0 \to \gamma \grav) &=& \frac{s_W^2}{s_W^2 + c_W^2\big( 1 - \frac{m_Z^2}{m^2_{\widetilde W}} \big)^4 }, \nonumber \\
{\rm BR}(\widetilde W^0 \to Z \grav) &=& 1 - {\rm BR}(\widetilde W^0 \to \gamma \grav).
\end{eqnarray}
Compared to the bino-like neutralino, the branching ratio to the photon final state is suppressed by the weak mixing angle squared, $s_W^2$, and the $Z$-boson mode is  dominant for winos heavier than $m_{\none} \gtrsim 200$\,GeV. In the limit $m_Z/m_{\widetilde W} \to 0$, they approach ${\rm BR}(\widetilde W^0 \to \gamma \grav) \to s^2_W \simeq 0.23$ and ${\rm BR}(\widetilde W^0 \to Z \grav) \to s^2_W \simeq 0.77$.

The higgsino-like neutralino, $\widetilde h_1^0$, decays into $\widetilde G$ and a Higgs or $Z$ boson. The branching ratios are calculated with $N_{13} = - N_{14} = 1/\sqrt{2}$ and $N_{11} = N_{12} = 0$ following Eq.~\eqref{eq:br_neut}. It is easy to see that ${\rm BR}(\widetilde h_1^0 \to \gamma \widetilde G) = 0$ and
\begin{eqnarray}
\frac{\Gamma(\widetilde h_1^0 \to Z \widetilde G)}{\Gamma(\widetilde h_1^0 \to h \widetilde G)} \; \simeq \; \frac{ 
|c_\beta + s_{\beta}|^2 \Big( 1 - m^2_Z/m^2_{\widetilde h_1^0} \Big)^4 }
{|c_\beta - s_{\beta}|^2 \Big( 1 - m^2_h/m^2_{\widetilde h_1^0} \Big)^4  } \,.
\end{eqnarray}
In the large $\tan \beta$ and heavy $\widetilde h_1^0$ limit, these mode will have the equal branching ratio of 50\%, though the $\widetilde h_1^0 \to Z \widetilde G$ mode is generally favoured due to the difference in phase-space and $\tan\beta$ effect.

\begin{figure}[t]
    \begin{minipage}{0.32\hsize}
        \begin{center}
        \includegraphics[width=5cm]{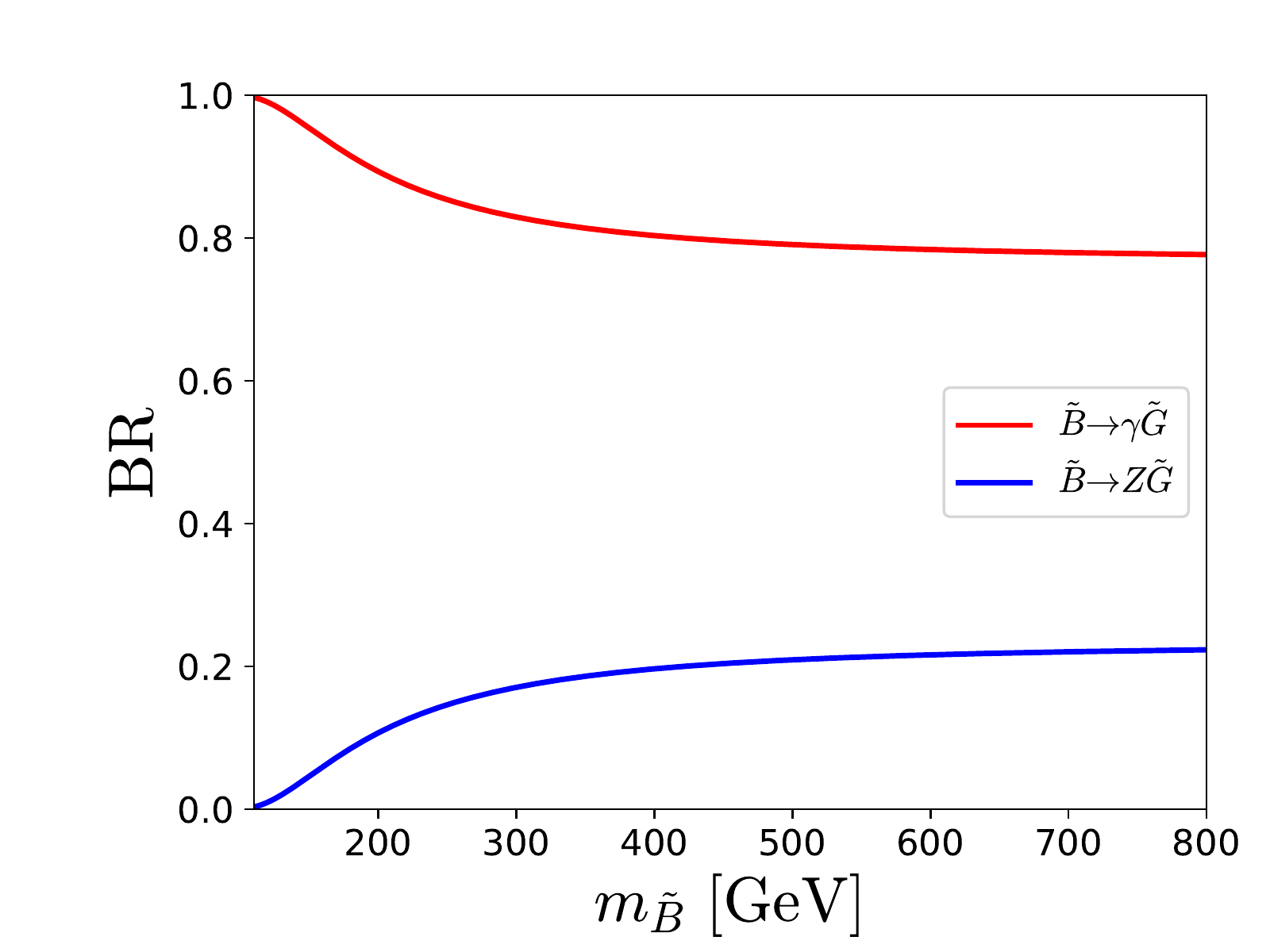}
        \end{center}
    \end{minipage}
    \begin{minipage}{0.32\hsize}
        \begin{center}
        \includegraphics[width=5cm]{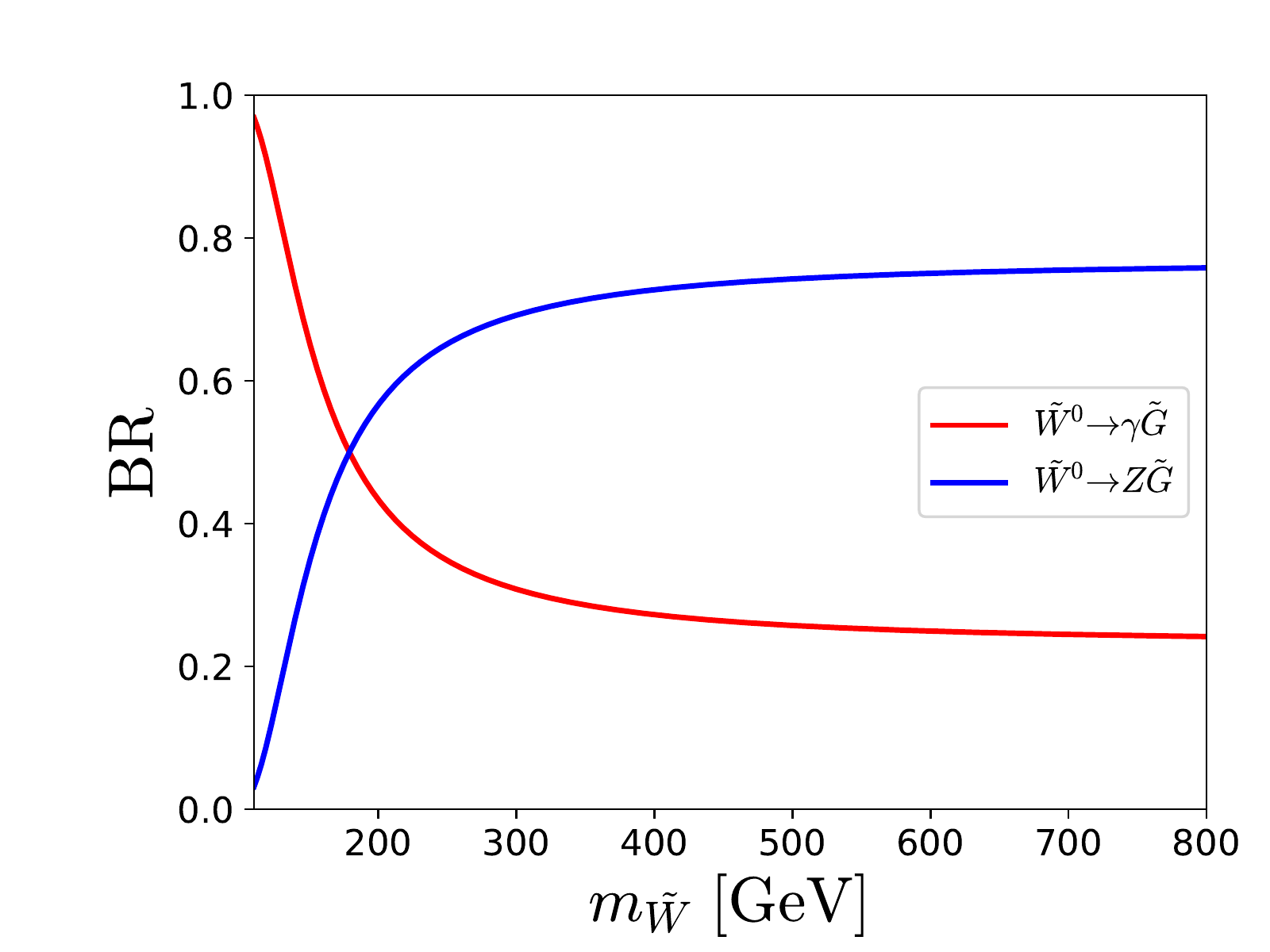}
        \end{center}
    \end{minipage}
    \begin{minipage}{0.32\hsize}
        \begin{center}
        \includegraphics[width=5cm]{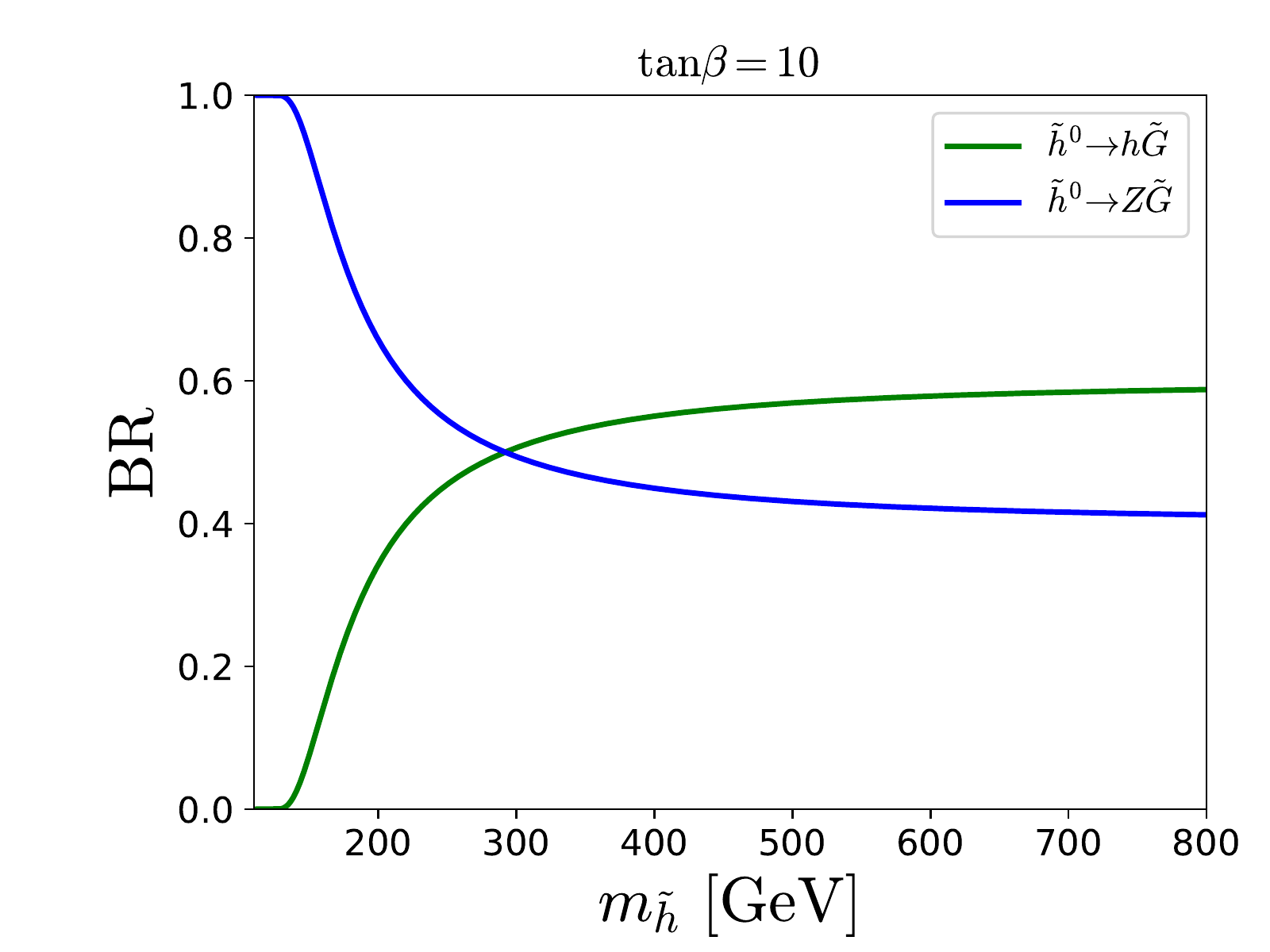}
        \end{center}
    \end{minipage}    
    \caption{
    \small Branching ratios to gravitino.
    \label{fig:brgravitino}    
    }
\end{figure}

In Figure~\ref{fig:brgravitino} we show the branching ratios of different classes of the NLSP. For binos the dominant decay mode is to the photon, regardless of its mass. For light winos photonic decay mode dominate as well, however for the heavier winos the dominant decay mode is to $Z$ boson. Finally higgsinos decay either to the Higgs boson or $Z$ boson, and for higgsinos heavier than 200~GeV either decay mode has a similar share.   

\subsection{Naturalness}

One of the motivations for a light gravitino is to relax the apparent fine-tuning in the Higgs sector. In the leading-log approximation, the Higgs mass-squared parameters  in a moderate or large $\tan\beta$ regime are roughly given by
\begin{equation}
m^2_h ~\sim~ 
\underbrace{ (m^2_{H_u} + |\mu|^2) }_{\rm tree}
~-~ 
\underbrace{ \frac{3 y_t^2}{8 \pi^2} m^2_{\rm stop} \log \frac{\Lambda}{Q} }_{\rm 1-loop}
~-~ 
\underbrace{ \frac{g_3^2 y_t^2}{4 \pi^4} |M_3|^2 \Big( \log \frac{\Lambda}{Q} \Big)^2 }_{\rm 2-loop} \,,
\label{eq:tuning}
\end{equation} 
where $Q$ is the electroweak symmetry breaking scale and $\Lambda$ is the messenger scale of SUSY breaking. The first, second and third terms come from tree, one-loop and two-loop level, and are sensitive to the higgsino mass ($\mu$), the stop mass ($m_{\rm stop}$) and the gluino mass ($M_3$), respectively. In our analysis we take all the mass parameters real and positive for simplicity. It is clear that the small $\mu$ is crucial for naturalness. From the above formula it is also evident that the second and third terms can be made not-too-large by taking $\Lambda$ to be small. The right panel of Fig.~\ref{fig:lifetime} shows the region of ($m_{\widetilde \chi_1^0}$, $\Lambda$) that is consistent with the prompt decay requirement $c \tau_{\none} \lesssim 1$\,mm. One can see that our prompt decay requirement is consistent with the region $\Lambda \lesssim $\,100--1000\,TeV, where the fine-tuning can be largely relaxed for given $\mu$, $m_{\rm stop}$ and $M_3$. 

In particular, for a meaningful estimation of the contributions from each of the three terms in Eq.~\eqref{eq:tuning} we have to specify the value of $\mu$. Therefore, we will discuss the impact of the gravitino LSP on the fine-tuning problem only for the higgsino-like neutralino case. In the other two cases, we have assumed that higgsinos are irrelevant for the collider signatures which, in practice, means that they are heavier than stops and gluinos. The limits obtained for stops and gluinos depend on this assumption. The first term in Eq.~\eqref{eq:tuning} is then the most important one irrespectively of the value of $\Lambda$.\footnote{One can of course relax this assumption and assume that higgsino is just heavier that the NLSP. The analysis  has then to be repeated with more signatures  taken into account and would lead to slightly weaker bounds on the stop and gluino masses, as a function of the assumed higgsino mass. Given that the obtained bounds are similar for all three simplified models, the discussion of the fine tuning issue just for the third model illustrates well the difference between the neutralino and gravitino LSP scenarios.}

\section{Recasting LHC analyses} 
\label{sec:recasting}

We confront our simplified models with various ATLAS and CMS analyses searching for beyond the Standard Model. For each analysed model we generate a grid of points each described by masses of particles characteristic for the model. The branching ratios are calculated using {\tt SDecay}~\cite{Muhlleitner:2003vg}. The exclusion of each point is determined using {\tt CheckMATE-2.0.26} \cite{Drees:2013wra}.

The events are generated within the {\tt CheckMATE} framework using {\tt Pythia 8} \cite{Sjostrand:2014zea} including parton shower and hadronization. The signal is simulated separately for the coloured particle production and electroweak particle production. The cross sections are rescaled to the next-to-the-leading-log (NLL) accuracy computed with {\tt NLL-fast} \cite{nllfast} and {\tt Resummino}~\cite{Fuks:2013vua} for coloured and electroweak particles, respectively.  The fast detector simulation is carried out using {\tt Delphes 3} \cite{deFavereau:2013fsa}. Finally, {\tt CheckMATE} emulates the various ATLAS and CMS analyses, estimating the signal efficiency  and confronting the signal yield with the model-independent 95\,\% CL upper limit for each signal region. In order to see the impact of individual analysis, we do not combine multiple analyses, but rather show the exclusion limit obtained from single analysis. 

The analyses used in this study are summarised in Table \ref{tb:analyses}. The above Monte Carlo chain is applied to the all analyses except for \verb|cms_1801_03957| \cite{1801_03957}. In \verb|cms_1801_03957|, the dedicated analysis has been carried out for the production of charginos and neutralinos decaying to a massless gravitino and a $Z$-boson or Higgs boson. The limit has been derived as a function of ${\rm BR}(\tilde h^0 \to h \tilde G)$  with ${\rm BR}(\tilde h^0 \to h \tilde G) + {\rm BR}(\tilde h^0 \to Z \tilde G) = 1$. The limit is strongest ($m_{\tilde h} \gtrsim 750$\,GeV) for ${\rm BR}(\tilde h^0 \to h \tilde G) = 1$, whilst most conservative ($m_{\tilde h} \gtrsim 650$\,GeV) for ${\rm BR}(\tilde h^0 \to h \tilde G) \sim 0.5$. Guided by our considerations in the previous section, see Figure~\ref{fig:brgravitino}, we impose the limit $m_{\tilde h} \gtrsim 650$\,GeV on our higgsino-gravitino simplified models.

\begin{table}[t!]
\begin{center}
  \begin{tabular}{|c|c|c|c|} \hline
  \texttt{CheckMATE} identifier & Description &  L\,[fb$^{-1}$] & ref.  \\ \hline \hline
  \verb|atlas_1709_04183|  & $\tilde t$, $b$-jets + $E_T^{\rm miss}$  &  36.1  & \cite{1709_04183}  \\ \hline
  \verb|atlas_1712_02332|  & $\tilde q$, $\tilde g$, jets + $E_T^{\rm miss}$  &  36.1  &  \cite{1712_02332}  \\  \hline
  \verb|atlas_1710_11412|  & dark matter with $t$ or $b$, $b$-jets, leptons  &  36.1  &  \cite{1710_11412}  \\  \hline
  \verb|atlas_1802_03158|  &  GMSB, $\gamma( \ge 1) + E_T^{\rm miss}$  &   36.1  &  \cite{1802_03158}  \\  \hline
  \verb|atlas_conf_2017_019|  & $\tilde t$, $Z$ or $h$ + $E_T^{\rm miss}$  &  36.1  &  \cite{ATLAS-CONF-2017-019} \\  \hline
  \verb|cms_1801_03957|  & electroweak, diboson final states ($W, Z, h$) &  35.9  &  \cite{1801_03957}  \\  \hline
  \verb|cms_sus_16_046|  & GMSB, $\gamma( \ge 1) + E_T^{\rm miss}$   &  35.9  &  \cite{CMS-PAS-SUS-16-046}  \\  \hline
  \verb|atlas_conf_2018_041|  &  $\tilde g$, $b$-jets + $E_T^{\rm miss}$  &  79.9 & \cite{ATLAS-CONF-2018-041}  \\ \hline
  \end{tabular}
  \caption{\small 
  The analyses used in this paper.
  \label{tb:analyses}}
\end{center}  
\end{table}

As we shall see in the following sections, apart from the already mentioned \verb|cms_1801_03957|, three other searches included in the current analysis will define exclusion limits. The first one is  \verb|atlas_1709_04183| which is the search for direct production of top squarks. The analysis searches for final states with several jets and at least one $b$-jet accompanied by large missing energy. The events with identified leptons are vetoed. It takes into account 36.1 fb$^{-1}$ of data collected in years 2015--2016. The second analysis is \verb|atlas_conf_2018_041| which is a search for gluino production decaying via third generation squarks. It looks for the final states with many (at least three) $b$-jets and large missing energy with or without leptons. It takes into account 79.9 fb$^{-1}$ of data collected in years 2015--2017, therefore one can expect that it provides very strong limits compared to other analyses. Finally the \verb|atlas_1802_03158| analysis searches for the production of electroweakinos, squarks and gluinos that subsequently decay to photons and gravitinos. The final state signature is in this case at least one isolated photon and large missing energy. 

\section{Results: the LHC limits}

Our presentation of the exclusion limits is organised in the following way. For each of the three models: stop, gluino and stop-gluino we dedicate a separate subsection. There we specify to models for which the strongly produced particles are accompanied by different classes of electroweakinos: bino, wino or higgsino, and finally we compare exclusions for the case with the electroweakino or gravitino LSP.   

\subsection{Stop simplified model \label{sec:stop}}

We start the analysis of the LHC constraints by looking at the simplified model with stops and electroweakinos. The pattern of stop decays will crucially depend on the nature of electroweakinos. Gluinos are assumed to be heavy, $m_{\widetilde{g}} \gtrsim 2.5$~TeV , which is well above the current limits.

In the simplest scenario, when $m_{\widetilde B} \ll m_{\widetilde W}, m_{\widetilde h}$, the lightest neutralino is predominantly composed of the bino. The only available decay mode is:
\beq
\textrm{stop-bino}:~~~
\tone \to t \none~~~~(\textrm{BR}=100\%),
\label{eq:stop-bino-decay}
\eeq
provided $m_{\tilde t_1} > m_t + m_{\tilde \chi_1^0}$. In the following analysis we assume that this relation is satisfied. Otherwise, $\tilde t_1$ may decay into $b W \tilde \chi_1^0$, $b jj \tilde \chi_1^0$ or $c \tilde \chi_1^0$ depending on the mass spectrum and the parameters.

In the wino-like neutralino scenario, the lightest stop will decay though its $\tilde t_L$ component into the winos. There are two possible decay modes and in the limit of heavy $\tone$, we have,
\begin{eqnarray}
\textrm{stop-wino}: && \tone \to b \widetilde W^+ ~~~~ (\textrm{BR} \simeq 2/3) \nonumber \\
&& \tone \to t \widetilde W^0 ~~~~ (\textrm{BR}  \simeq 1/3).
\end{eqnarray}
For smaller stop mass, the phase-space factor becomes important, which further favours the $\tone \to b \widetilde W^+$ mode. In particular, for $m_{\tone} < m_{\widetilde W} + m_t$, the the top-quark decay mode vanishes and $\textrm{BR} (\tone \to b \widetilde W^+) = 100\%$.

In the higgsino scenario there are three competing stop decay modes:
\begin{eqnarray}
\textrm{stop-higgsino}: && \tone \to t \widetilde h_1^0 ~~~~ (\textrm{BR} \simeq 25\%) \nonumber \\
&& \tone \to b \widetilde h^+ ~~~~ (\textrm{BR}  \simeq 50\%) \nonumber \\
&& \tone \to t \widetilde h_2^0 ~~~~ (\textrm{BR} \simeq 25\%)
\end{eqnarray}
where the branching ratios are in the $m_{\tone} \gg m_{\widetilde h}$ limit. On the other hand, for lighter stops the $\tone \to b \widetilde h^+$ mode is preferred due to larger phase-space. In particular $\textrm{BR} (\tone \to b \widetilde h^+) = 100\%$ for $m_{\tone} < m_{\widetilde h} + m_t$. In our Monte Carlo simulation the branching ratios for different parameter points are obtained by {\tt SDecay}~\cite{Muhlleitner:2003vg}.

\begin{figure}[th!]
    \begin{minipage}{0.5\hsize}
        \begin{center}
        \includegraphics[width=8cm]{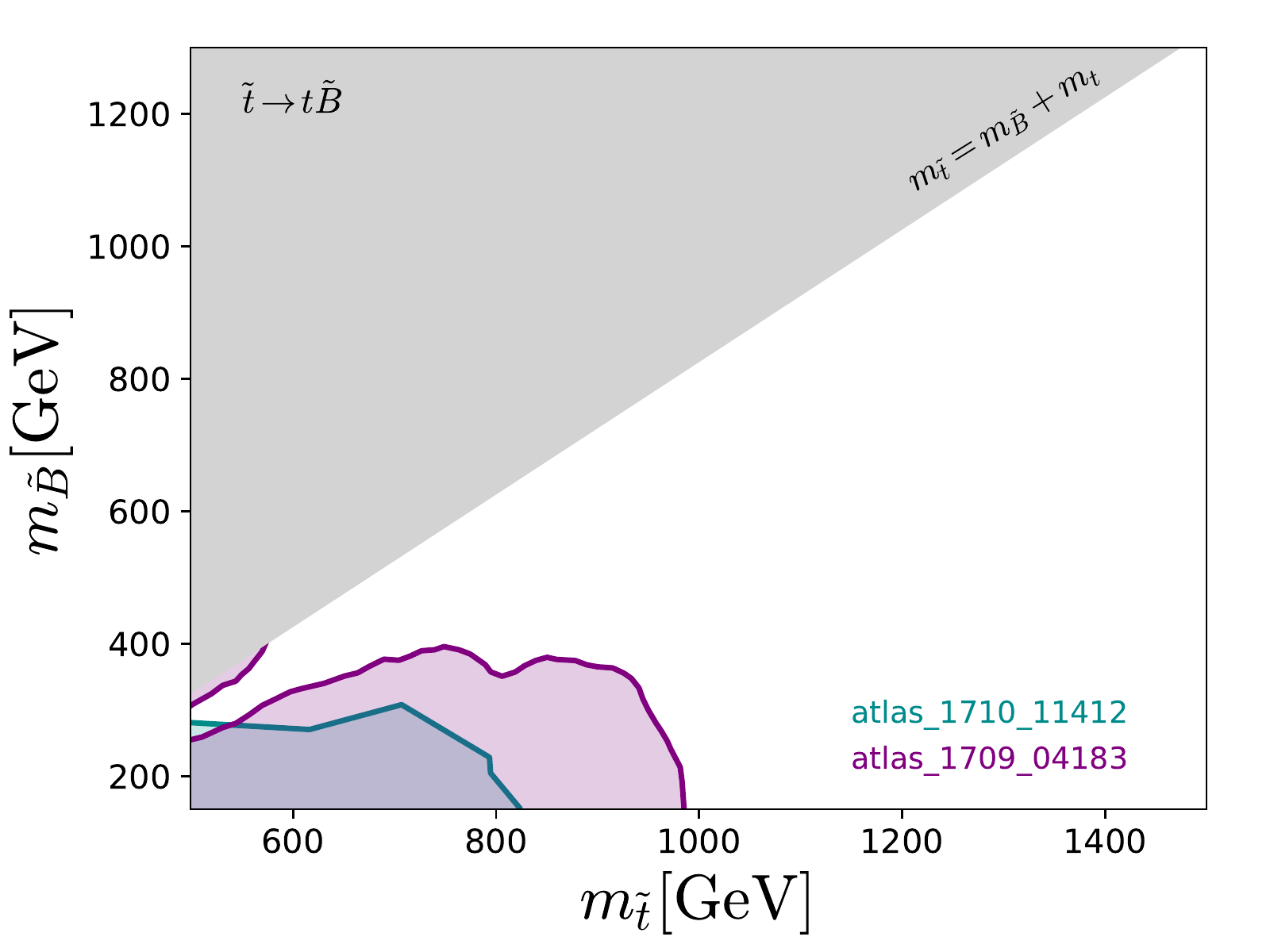}
        \end{center}
    \end{minipage}
    \begin{minipage}{0.5\hsize}
        \begin{center}
        \includegraphics[width=8cm]{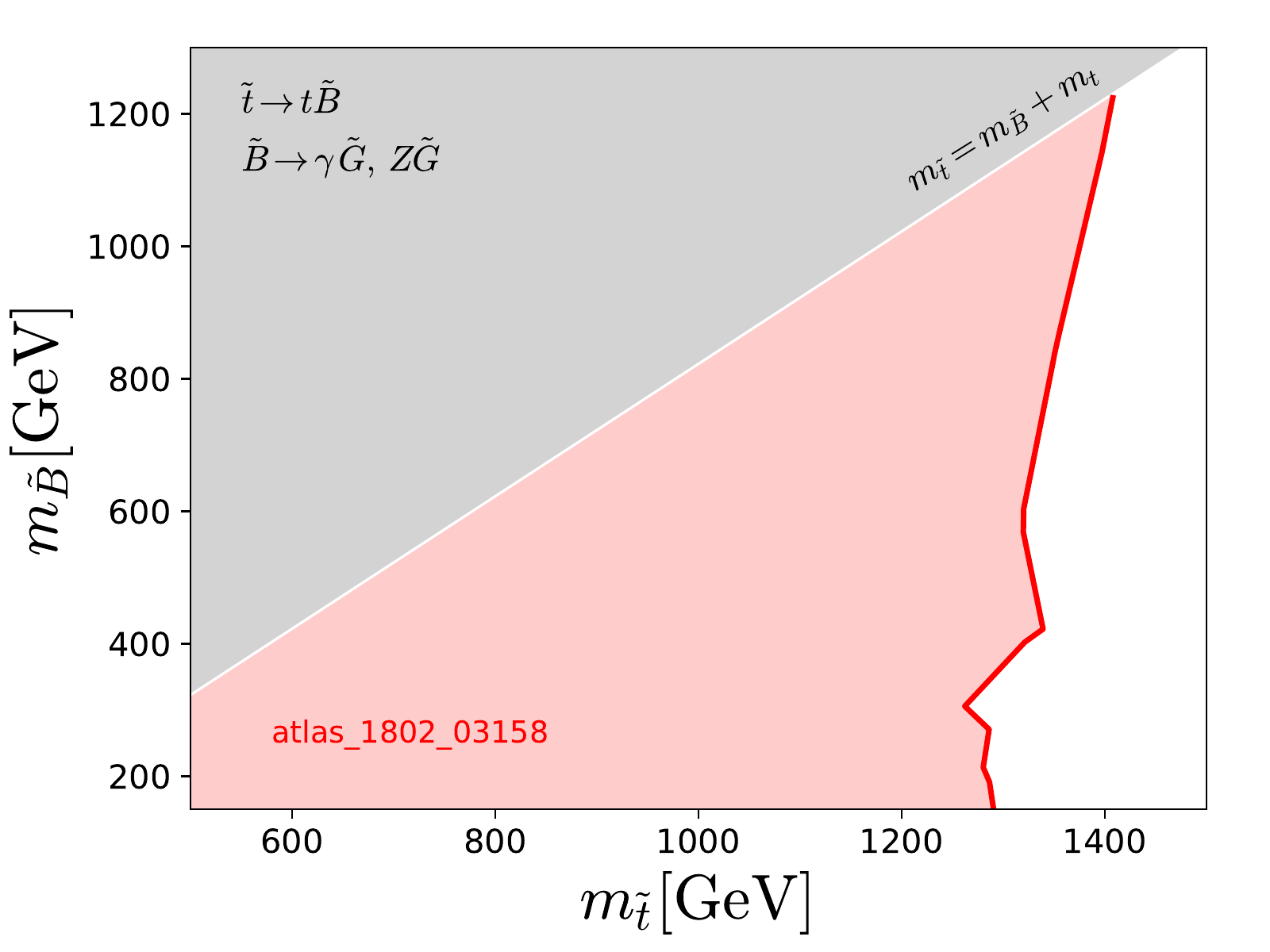}
        \end{center}
    \end{minipage}
    \begin{minipage}{0.5\hsize}
        \begin{center}
        \includegraphics[width=8cm]{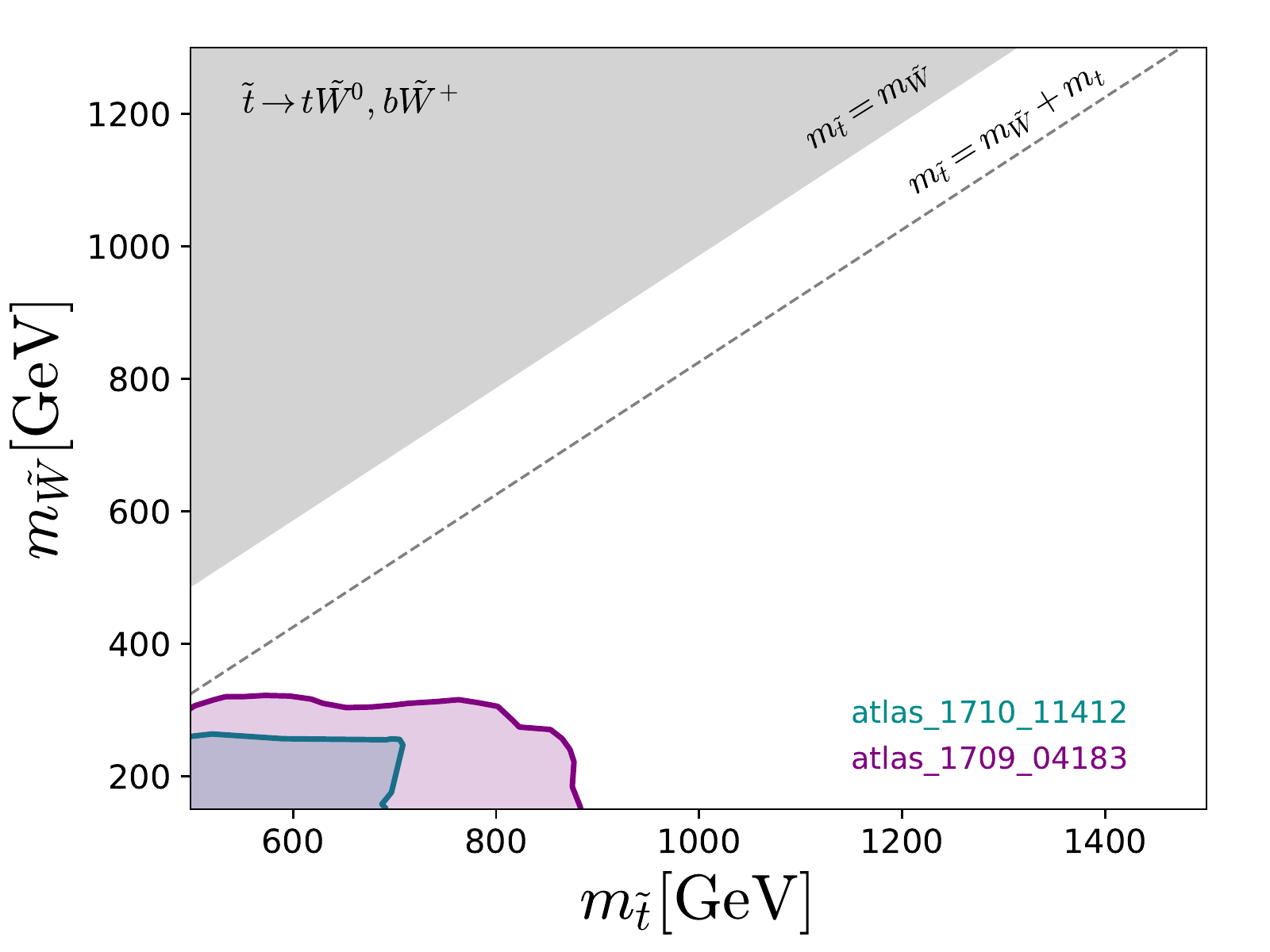}
        \end{center}
    \end{minipage}
    \begin{minipage}{0.5\hsize}
        \begin{center}
        \includegraphics[width=8cm]{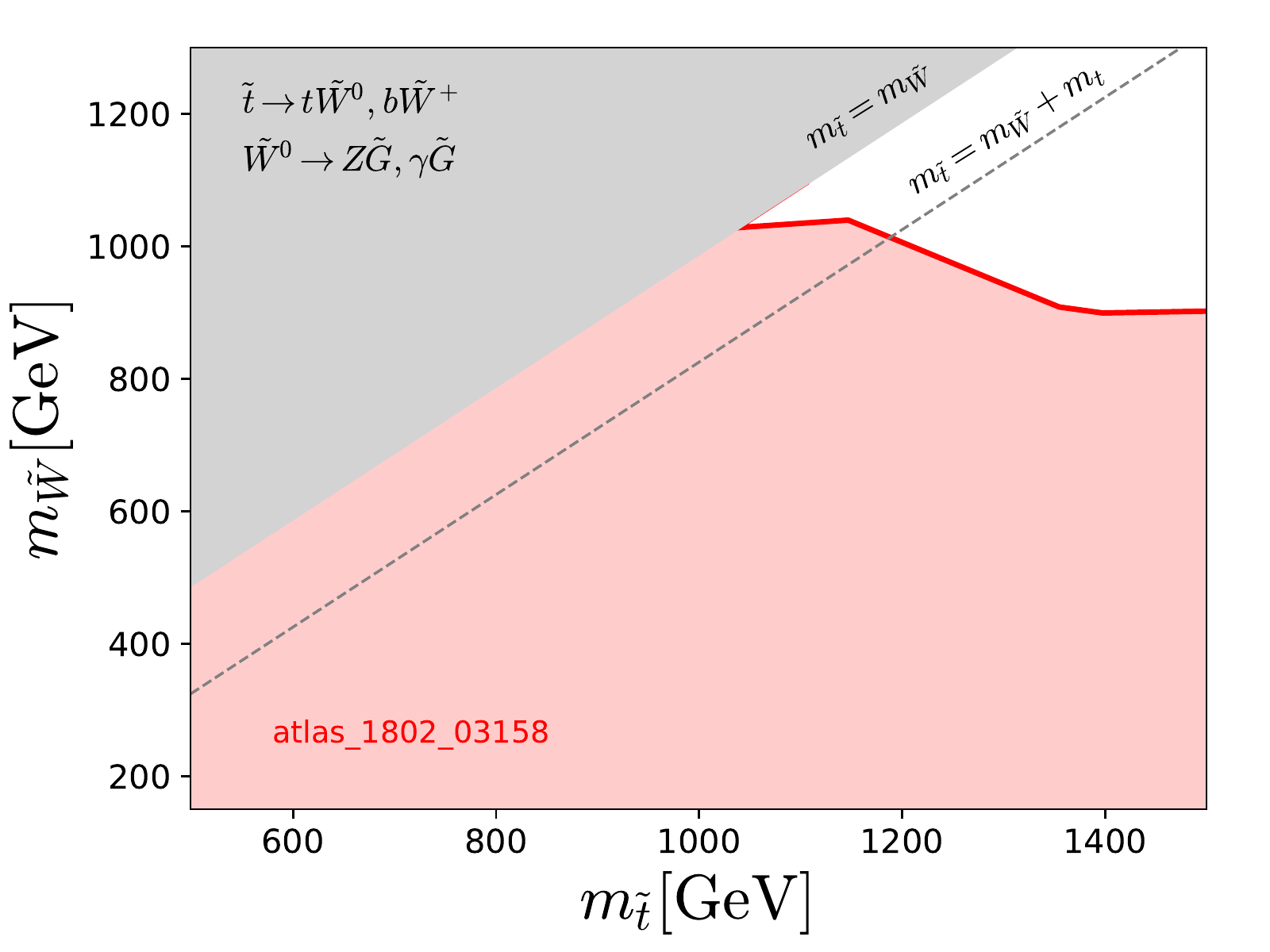}
        \end{center}
    \end{minipage}
    \begin{minipage}{0.5\hsize}
        \begin{center}
        \includegraphics[width=8cm]{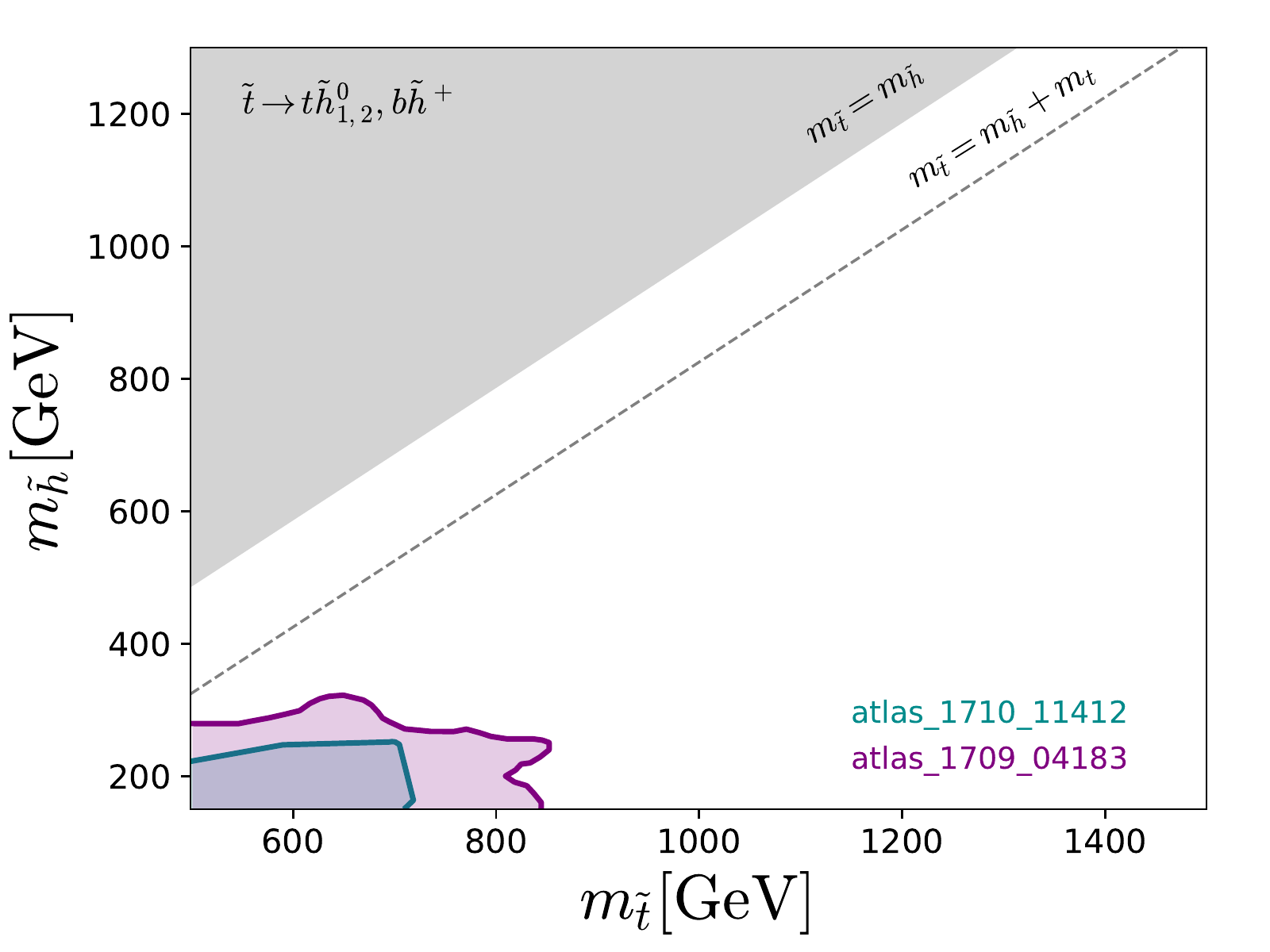}
        \end{center}
    \end{minipage}
    \begin{minipage}{0.5\hsize}
        \begin{center}
        \includegraphics[width=8cm]{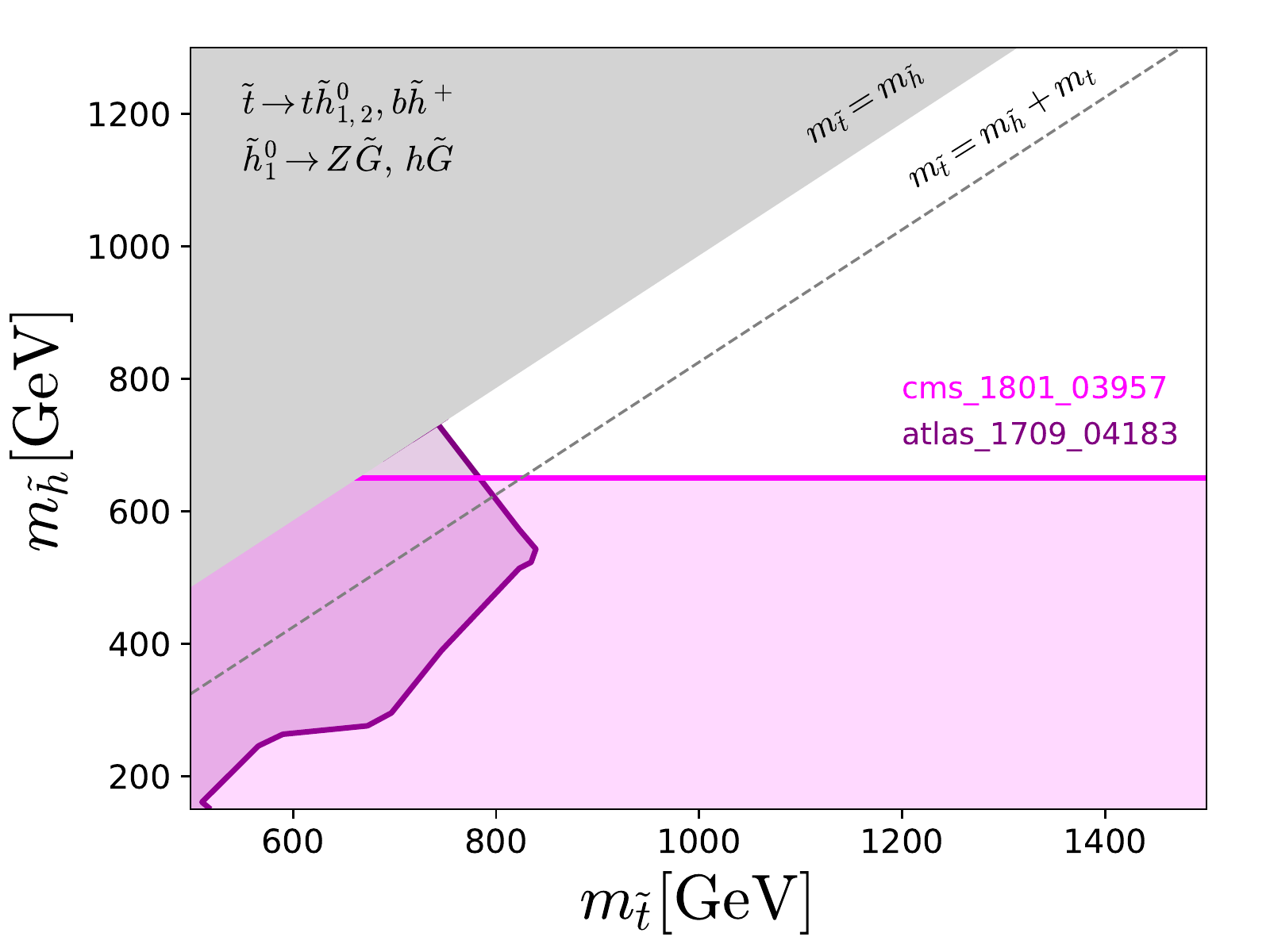}
        \end{center}
    \end{minipage}
    \caption{
    \small The exclusion plots in the stop model, Sec.~\ref{sec:stop}. The left column with electroweakino as the LSP and the right column with gravitino LSP. From top to bottom: bino, wino and higgsino case. Only the most constraining analyses are shown.
    }
    \label{fig:stop}
\end{figure}

The LHC constraints on the stop simplified model are summarized in Fig.~\ref{fig:stop}, without gravitino (left column) and with gravitino (right column). For the model without gravitino the strongest constraint comes from the ATLAS search for direct stop production \verb|atlas_1709_04183| which targets stops decaying into top and the neutralino or into a bottom quark and chargino with subsequent decays via (off-shell) $W$'s. The limit is the strongest for the bino LSP and extends up to $m_{\widetilde{t}} \sim 1000$~GeV which is consistent with the ATLAS results. For the wino and higgsino case the limit is slightly weaker due to different competing decay modes and extends up to $m_{\widetilde{t}} = 850$--$900$~GeV.  

The exclusion limit is drastically changed when a light gravitino is added as the lightest supersymmetric particle. In this case, we have additional handles from the decay products of neutralinos at the expense of missing transverse energy. As a result, as can be seen below, the limits for the gravitino LSP models are much stronger than for the neutralino LSP models in general. In the NLSP bino case (top right plot in Fig.~\ref{fig:stop}), we find \verb|atlas_1802_03158| (GMSB;~ $\gamma( \ge 1) + E_T^{\rm miss}$) \cite{1802_03158} is by far the strongest, excluding $\widetilde{t}_1$ up to $m_{\widetilde{t}_1} \gtrsim 1250$ GeV for $m_{\widetilde B} \sim 200$ GeV and $m_{\widetilde{t}_1} \gtrsim 1400$ GeV for $m_{\widetilde B} \sim 1000$ GeV. This analysis have managed to reduce the Standard Model background  drastically by requiring at least one energetic photon in conjunction with large $E_T^{\rm miss}$. Unlike the previous case without gravitino, the limit is not weakened for heavier bino, because the photons produced from $\widetilde B$ decays are more energetic for larger $m_{\tilde B}$ in general. 

In the middle right panel we see the limit on the stop-wino-gravitino model. As in the bino-like $\none$ scenario with the light gravitino, \verb|atlas_1802_03158| is the most sensitive across the plane. The exclusion is driven by the electroweak production of charginos, $\widetilde{\chi}^+_1 \widetilde{\chi}^-_1$, chargino-neutralino pairs, $\widetilde{\chi}^\pm_1 \widetilde{\chi}^0_1$, and the stop production. Note that in the bino case, the contribution from bino production is negligible due to its very weak interaction and the only supersymmetric production is that of stops. This explains the striking difference between the shape of exclusion lines in both cases. In the wino case, the purely electroweak production gives rise to the horizontal exclusion line at $m_{\widetilde W} \sim 900$ GeV. For lighter stops due to the additional small contribution from stop production the limit goes as far as $m_{\widetilde W} \gtrsim 1050$~GeV.  Generally we find the limit, $m_{\tone} \gtrsim 1050$ GeV, which is slightly weaker than the corresponding limit in bino-like $\none$ scenario. This is because the branching ratio for $\none \to \gamma \widetilde G$  is suppressed by $s_W^2$ compared to the bino-like $\none$ case, therefore the contribution to the diphoton signal is reduced. We see that \verb|atlas_1802_03158| excludes winos below $\sim 900$ GeV almost model-independently, as long as a massless gravitino is present in the spectrum. 

The lower right panel shows the limit on the stop-higgsino-gravitino model. As can be seen, there is a strong constraint on the higgsino mass, $m_{\tilde h} \gtrsim 650$\,GeV, coming from \verb|cms_1801_03957| (electroweak; diboson final states [$W, Z, h$]) \cite{1801_03957}. This limit has been obtained in \verb|cms_1801_03957| by a dedicated analysis targeting pair productions of higgsino states decaying into a massless gravitino. For the higgsino heavier than $650$~GeV the relevant limit on the stop mass comes from \verb|atlas_1709_04183| ($\tilde t$;~ $b$-jets + $E_T^{\rm miss}$) \cite{1709_04183}, which exclude the stop up to $\sim 750$~GeV.

\subsection{Gluino simplified model\label{sec:gluino}}

The second type of the analysed simplified model assumes that the only SUSY particles accessible at the LHC are gluinos and electroweakinos. Stops are assumed to be lighter than other squarks but heavier than gluinos, so that the gluino decays via the off-shell top quarks in the three-body decay modes. The pattern of gluino decays will therefore follow the pattern of stop decays discussed in the previous section:
\begin{eqnarray}
 \textrm{gluino-bino}:&&~~~\gluino \to t \bar{t} \none~~~~(\textrm{BR}=100\%),\nonumber \\ 
 \textrm{gluino-wino}:&&~~~\gluino \to t \bar{t} \none~~~~(\textrm{BR}=67\%),\nonumber \\ 
                      &&~~~\gluino \to t \bar{b} \tilde{\chi}^-_1\ +\ \mathrm{c.c}~~~~(\textrm{BR}=33\%),\nonumber \\ 
 \textrm{gluino-higgsino}:&&~~~\gluino \to t \bar{t} \none~~~~(\textrm{BR}=25\%),\nonumber \\ 
                      &&~~~\gluino \to t \bar{t} \tilde{\chi}^0_2~~~~(\textrm{BR}=25\%),\nonumber \\ 
                      &&~~~\gluino \to t \bar{b} \tilde{\chi}^-_1\ +\ \mathrm{c.c}~~~~(\textrm{BR}=50\%).       \nonumber            
\end{eqnarray}
Note that for the wino and higgsino cases the above branching ratios assume that $m_{\gluino} \gg m_{\widetilde{\chi}}$. For smaller gluino mass, the $\gluino \to t b \tilde W^\pm$ mode becomes more favoured by the phase-space. In particular, $Br(\gluino \to t t \tilde W^0) = 0$ for $m_{\gluino} < 2 m_t + m_{\widetilde W}$. In our Monte Carlo simulation, we calculate the stop and gluino branching ratios using {\tt SDecay} \cite{Muhlleitner:2003vg}.

\begin{figure}[h!]
	\begin{minipage}{0.5\hsize}
		\begin{center}
		\includegraphics[width=8cm]{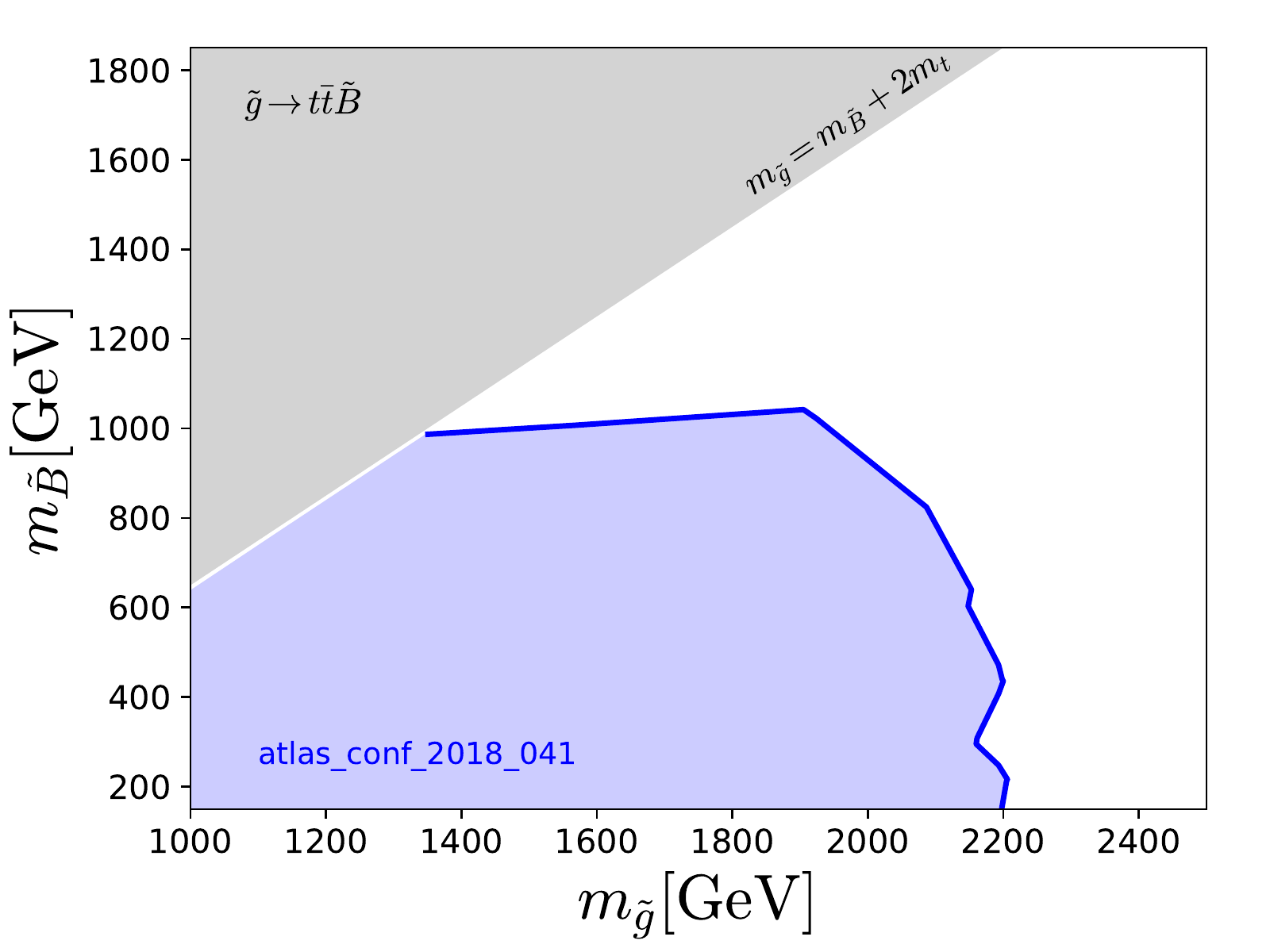}
		\end{center}
	\end{minipage}
	\begin{minipage}{0.5\hsize}
		\begin{center}
		\includegraphics[width=8cm]{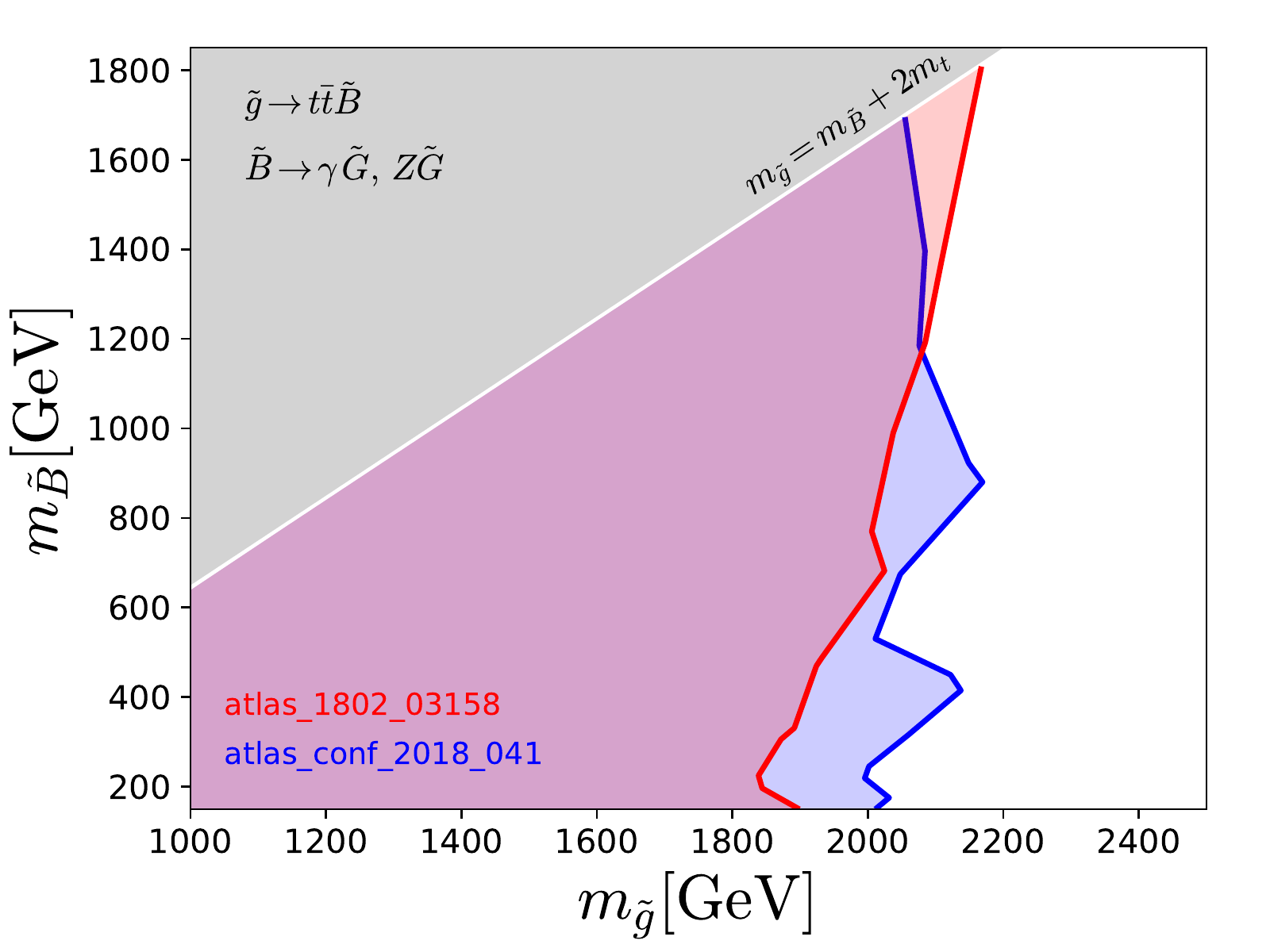}
		\end{center}
	\end{minipage}
    \begin{minipage}{0.5\hsize}
        \begin{center}
        \includegraphics[width=8cm]{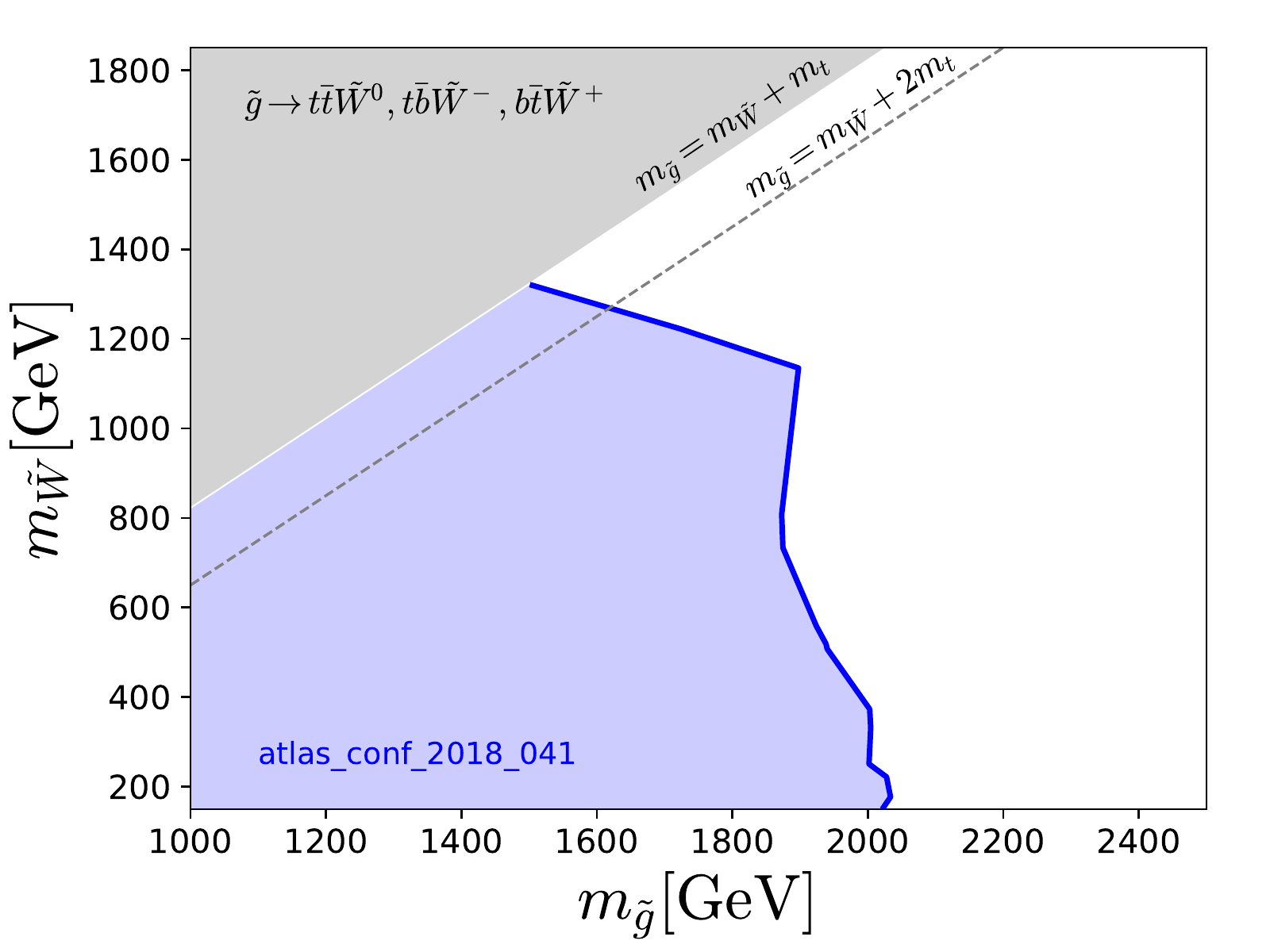}
        \end{center}
    \end{minipage}
    \begin{minipage}{0.5\hsize}
        \begin{center}
        \includegraphics[width=8cm]{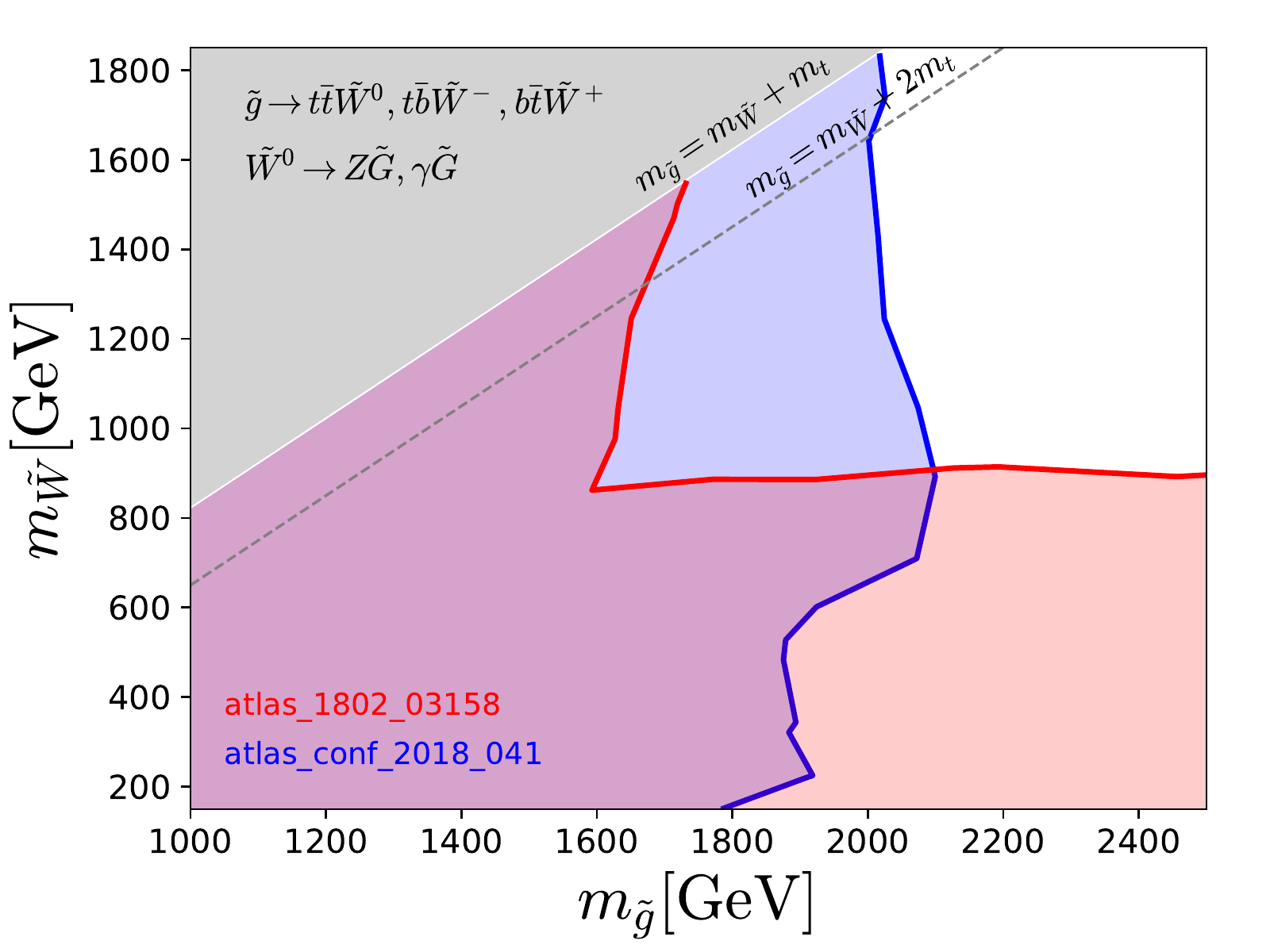}
        \end{center}
    \end{minipage}
    \begin{minipage}{0.5\hsize}
        \begin{center}
        \includegraphics[width=8cm]{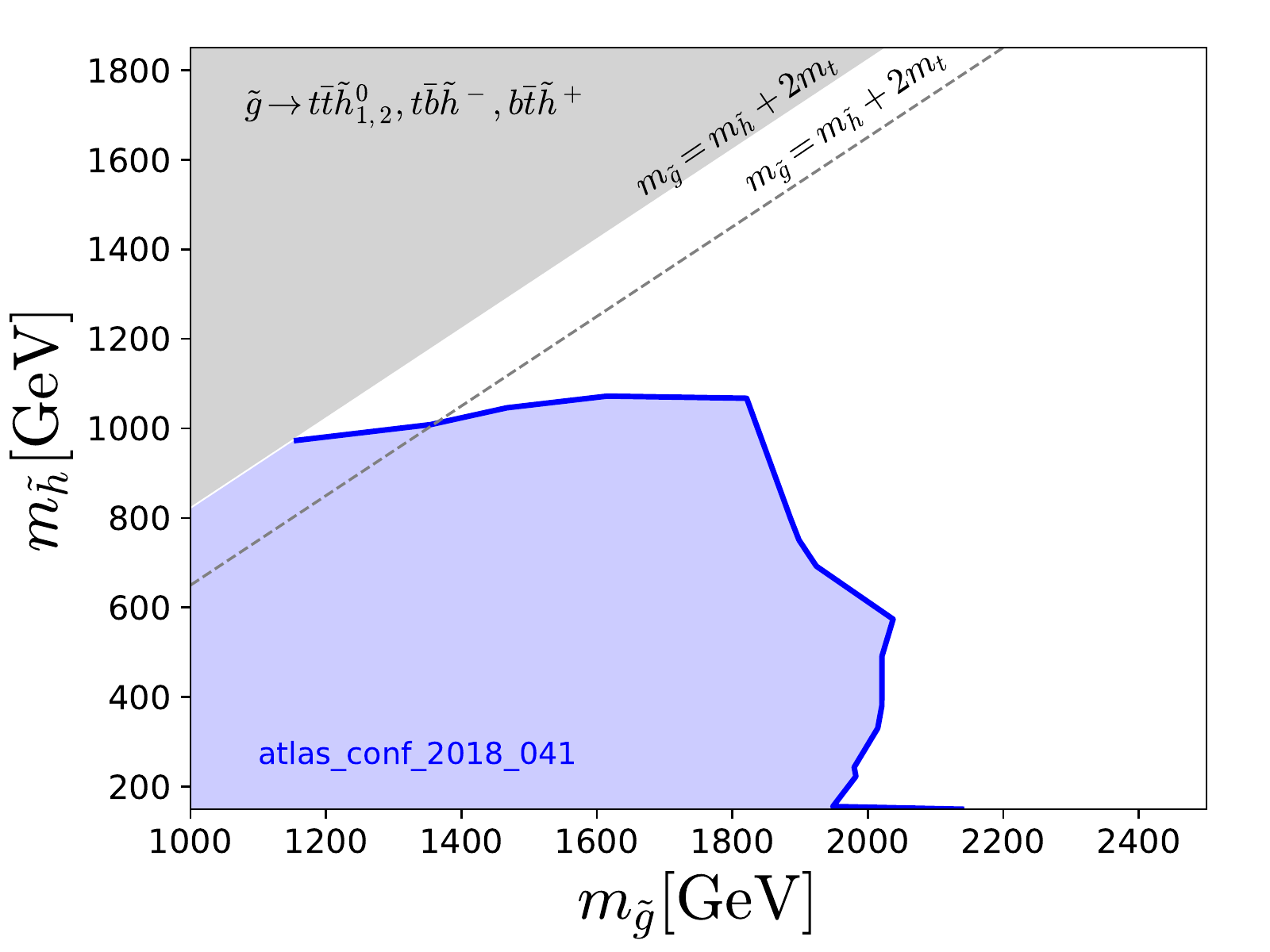}
        \end{center}
    \end{minipage}
    \begin{minipage}{0.5\hsize}
        \begin{center}
        \includegraphics[width=8cm]{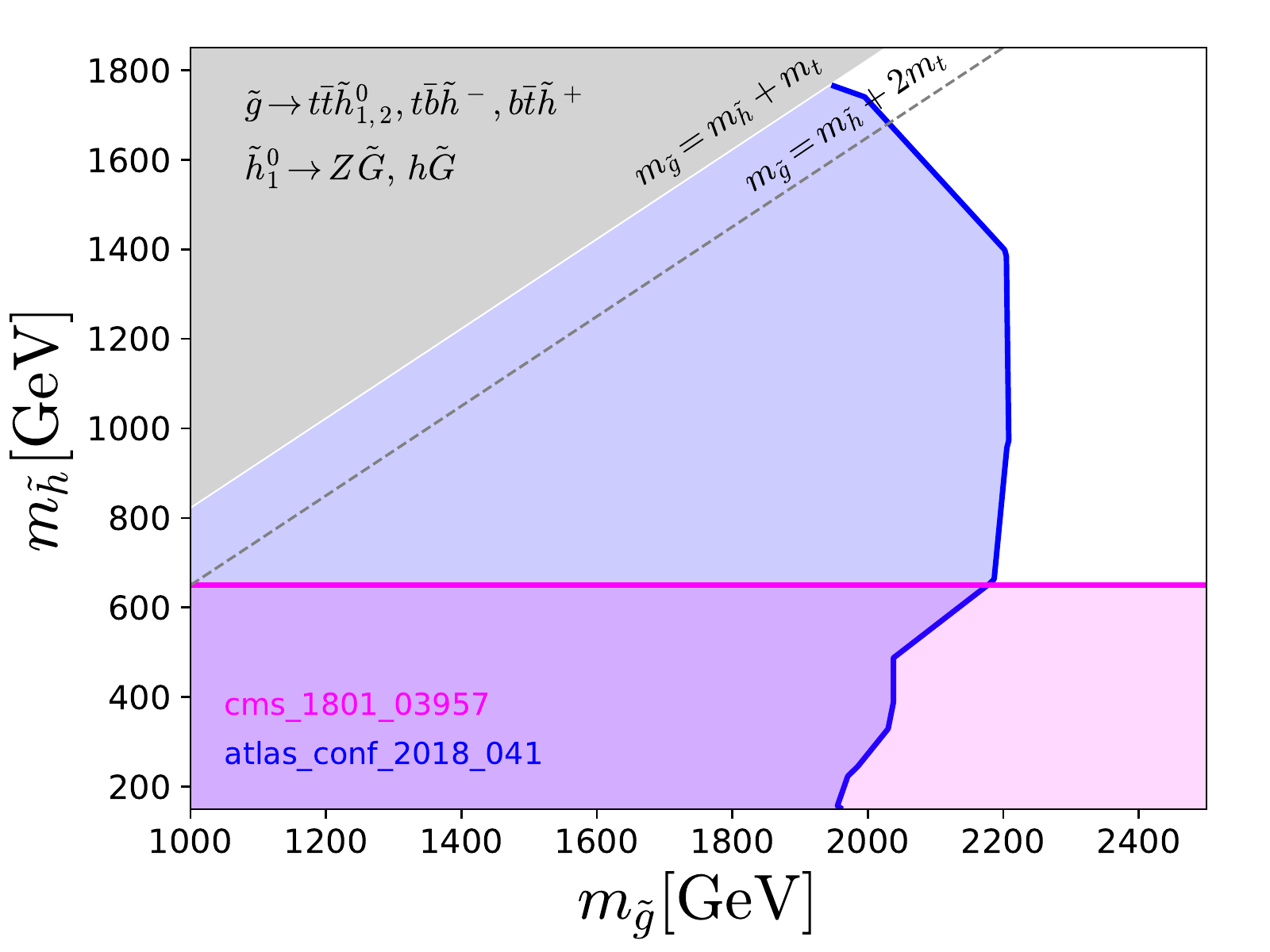}
        \end{center}
    \end{minipage}
    \caption{
    \small The exclusion plots in the gluino model, Sec.~\ref{sec:gluino}. The left column with electroweakino as the LSP and the right column with gravitino LSP. From top to bottom: bino, wino and higgsino case. Only the most constraining analyses are shown.
    }
    \label{fig:gluino}
\end{figure}

The LHC constraints on the gluino simplified model are summarized in Fig.~\ref{fig:gluino}, without gravitino (left column) and with gravitino (right column). For the model without gravitino the strongest constraint comes from the ATLAS search for direct stop production \verb|atlas_conf_2018_041|. It provides stricter limits than \verb|atlas_1709_04183| which considers stops produced in gluino decays. However, the latter search only takes into account pre-2018 results.

The limit is the strongest for the bino LSP and extends up to $m_{\tilde{g}} \sim 2200$~GeV. For the wino and higgsino case the limit is slightly weaker due to different competing decay modes and extends up to $m_{\tilde{g}} \sim 2000$~GeV. In either case the exclusion is up to the LSP mass of  $\sim 1000$~GeV for the bino and higgsino scenarios, while the limit extends to $\sim 1200$ GeV in the wino case. 

For the models with the gravitino LSP, the right column of Fig.~\ref{fig:gluino}, the shape of excluded region depends on the nature of the NLSP. In the bino case, two searches \verb|atlas_conf_2018_041| and \verb|atlas_1802_03158| provide comparable limits and gluino with masses up to 2000--2200~GeV are excluded, depending on the NLSP mass. While the \verb|atlas_1802_03158| search targets specifically the final states with energetic photons it is using just 36~fb$^{-1}$ of data. On the other hand \verb|atlas_conf_2018_041| does not take the advantage of the additional photons, but includes additional 44~fb$^{-1}$ of data from year 2017. 

The situation is different for the wino NLSP. As we have seen in the stop-wino-gravitino model, the light wino below $\sim 900$ GeV is excluded independently of the gluino mass by \verb|atlas_1802_03158|  exploiting the photon final state. The limit is again due to the electroweak production of chargino, $\widetilde{\chi}^+_1 \widetilde{\chi}^-_1$, and chargino-neutralino  pairs, $\widetilde{\chi}^\pm_1 \widetilde{\chi}^0_1$.  However, in the heavy wino region, $m_{\widetilde W} \gtrsim 900$ GeV, the most stringent constraint comes from \verb|atlas_conf_2018_041| exploiting the hadronic final state with multiple $b$-jets accompanied by missing energy. This limit only slightly depends on the wino mass, and exclude the gluino up to $\sim 2000$ GeV. It is easy to understand why the dedicated GMSB search \verb|atlas_1802_03158| gives a weaker exclusion for heavy winos looking at Fig.~\ref{fig:brgravitino}. In the wino scenario, the branching ratio ${\rm Br}(\widetilde W^0 \to \gamma \grav)$ becomes smaller than the $Z$ decay mode, contrary to the bino case.

Similarly to the stop simplified model discussed in the previous section, higgsinos with mass $m_{\widetilde h} \lesssim 650$\,GeV, are excluded by \verb|cms_1801_03957|. For heavier higgsinos, the strongest limit comes again from \verb|atlas_conf_2018_041|, extending up to 2200~GeV. The limit becomes slightly weaker for higgsinos heavier than $\sim 1500$ GeV.

\subsection{Stop-gluino simplified model\label{sec:stop-gluino}}

\begin{figure}[h!]
    \begin{minipage}{0.5\hsize}
        \begin{center}
        \includegraphics[width=8cm]{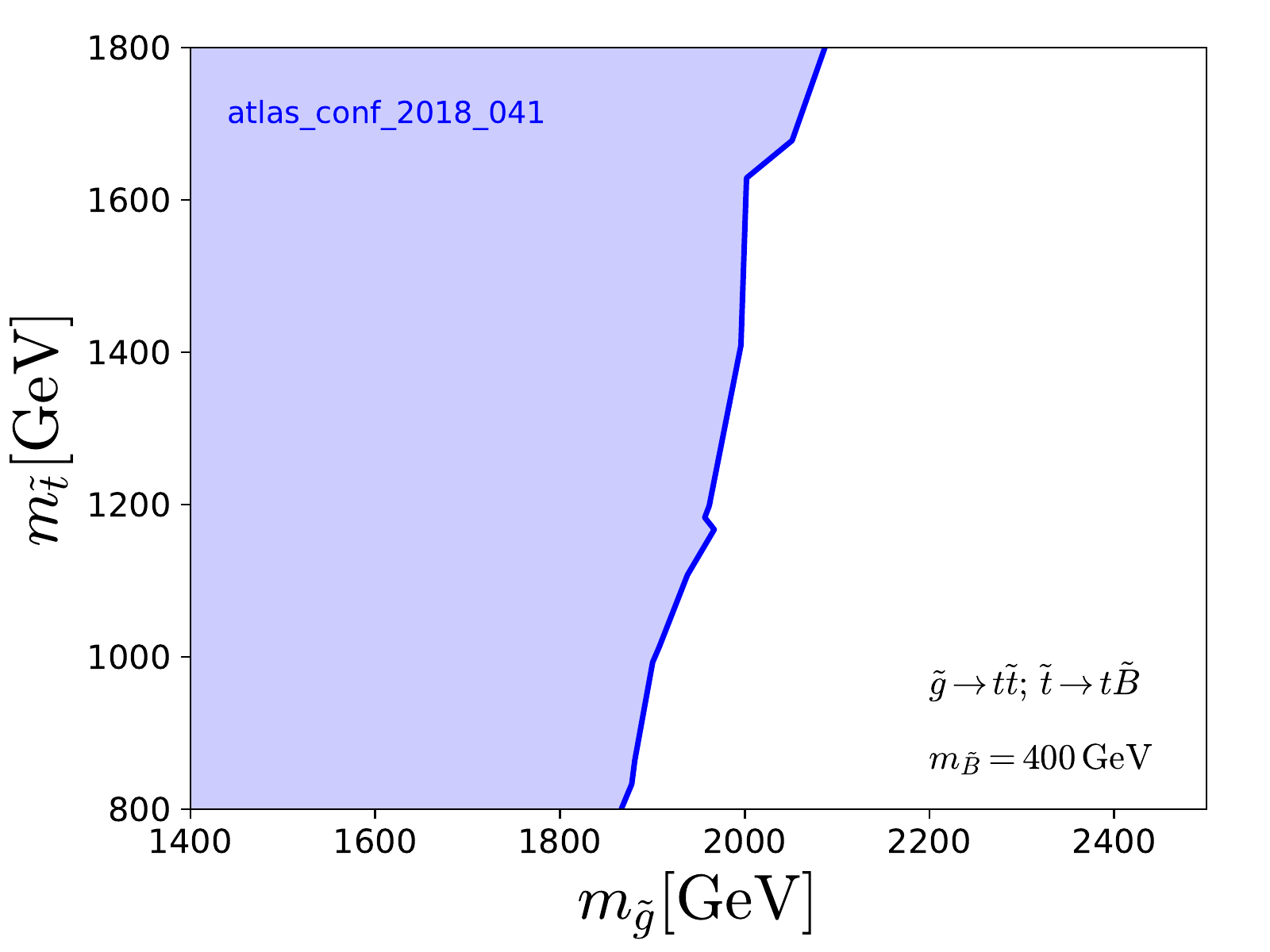}
        \end{center}
    \end{minipage}
    \begin{minipage}{0.5\hsize}
        \begin{center}
        \includegraphics[width=8cm]{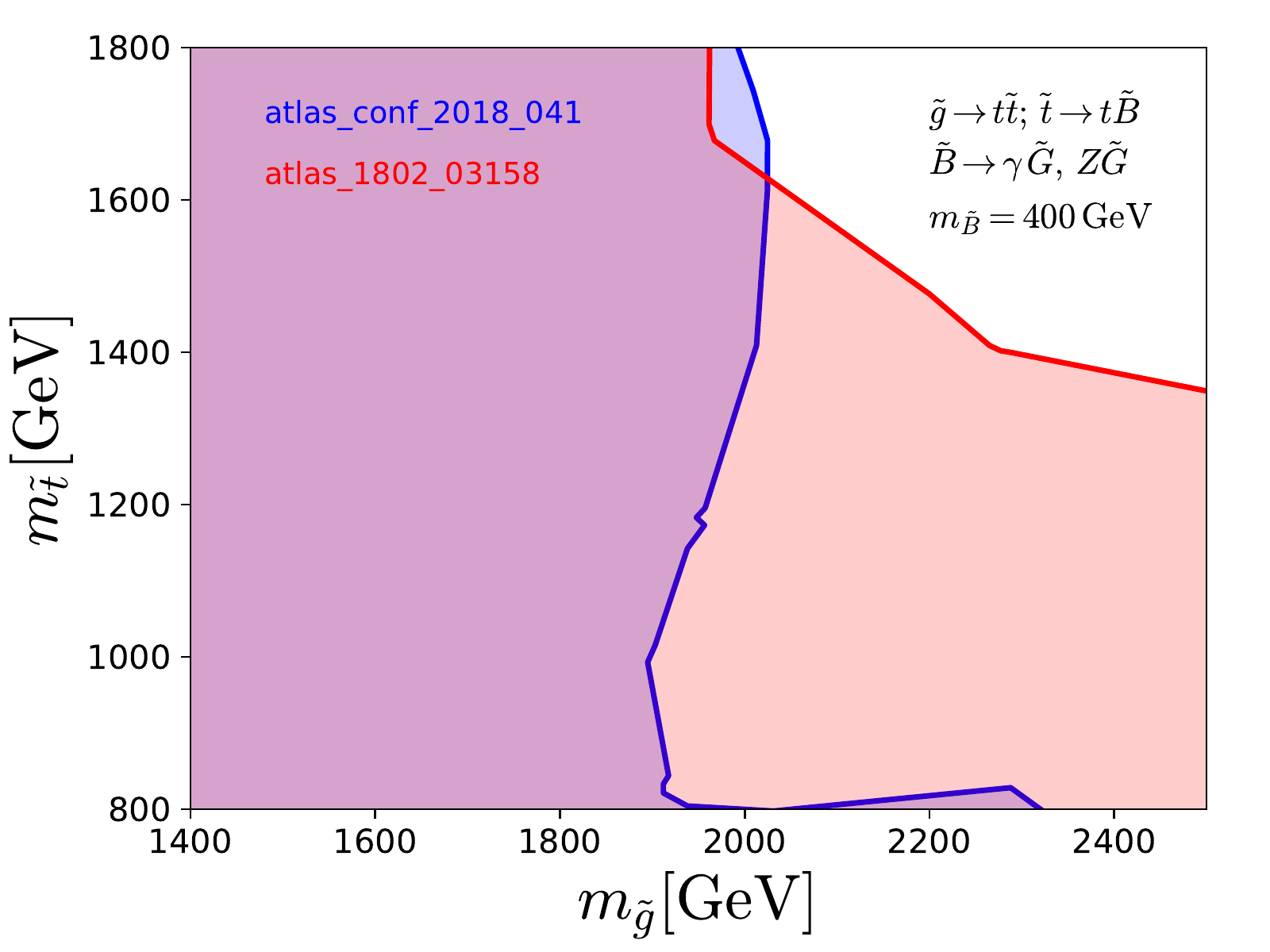}
        \end{center}
    \end{minipage}
    \begin{minipage}{0.5\hsize}
        \begin{center}
        \includegraphics[width=8cm]{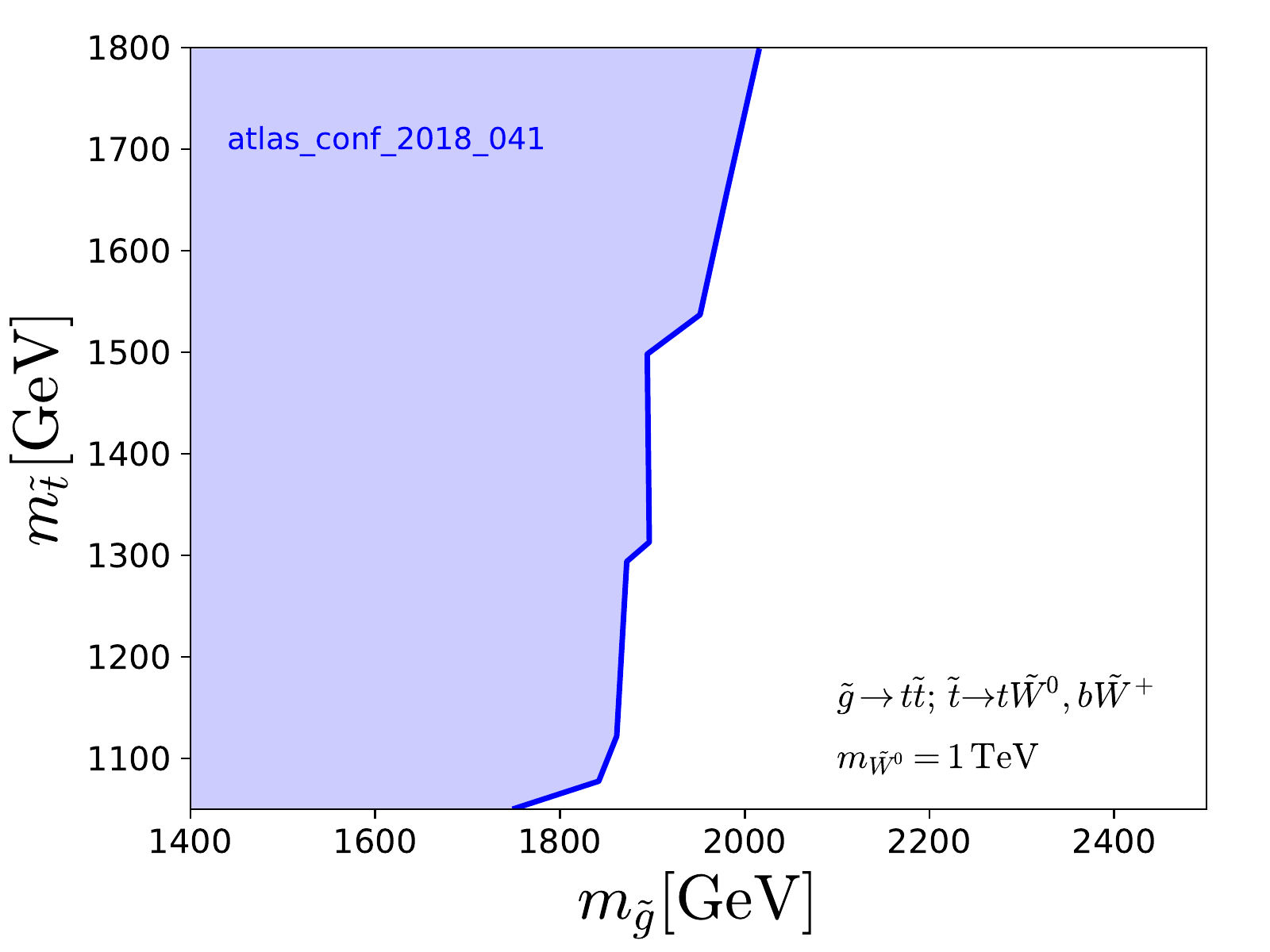}
        \end{center}
    \end{minipage}
    \begin{minipage}{0.5\hsize}
        \begin{center}
        \includegraphics[width=8cm]{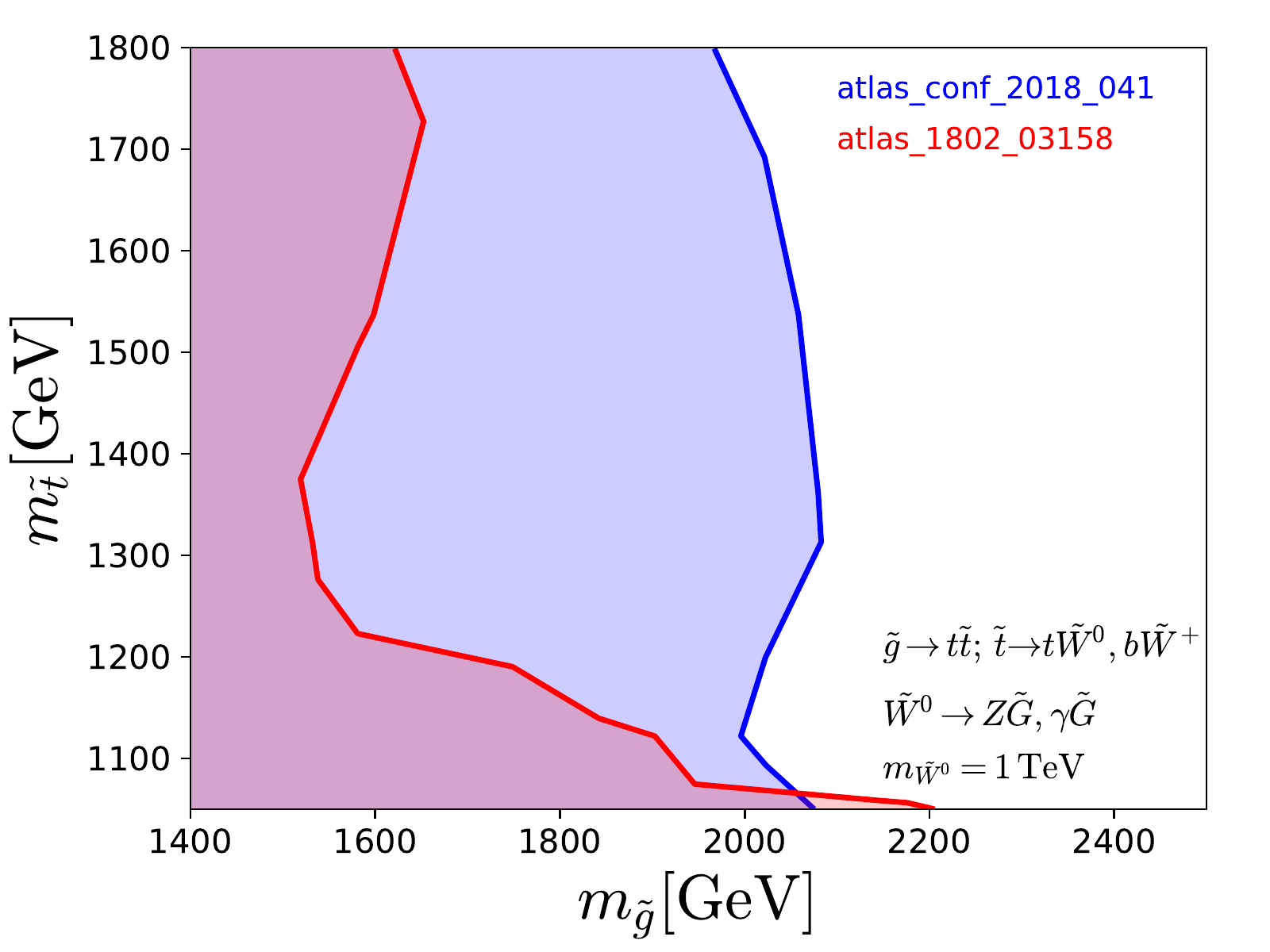}
        \end{center}
    \end{minipage}
    \begin{minipage}{0.5\hsize}
        \begin{center}
        \includegraphics[width=8cm]{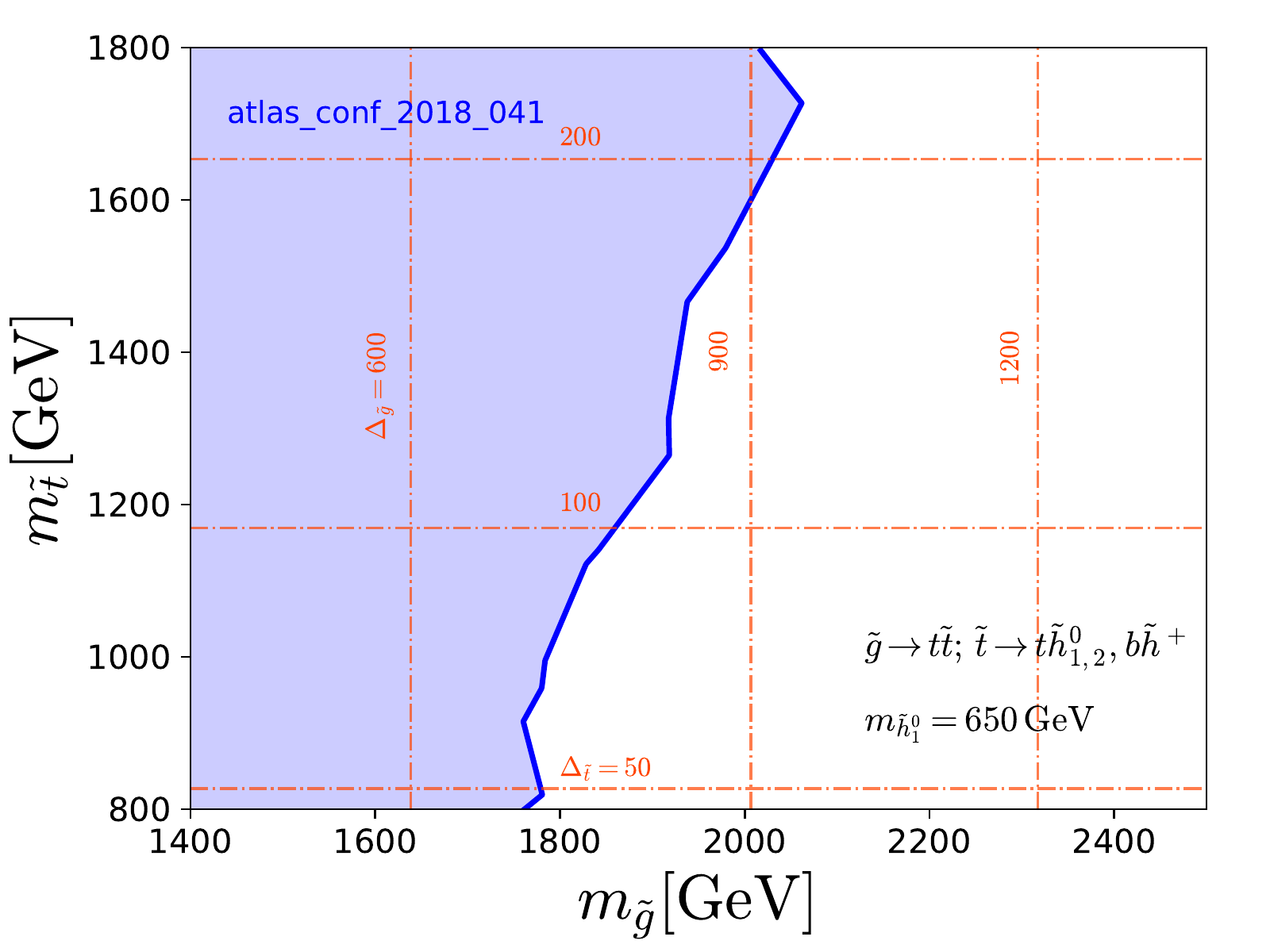}
        \end{center}
    \end{minipage}
    \begin{minipage}{0.5\hsize}
        \begin{center}
        \includegraphics[width=8cm]{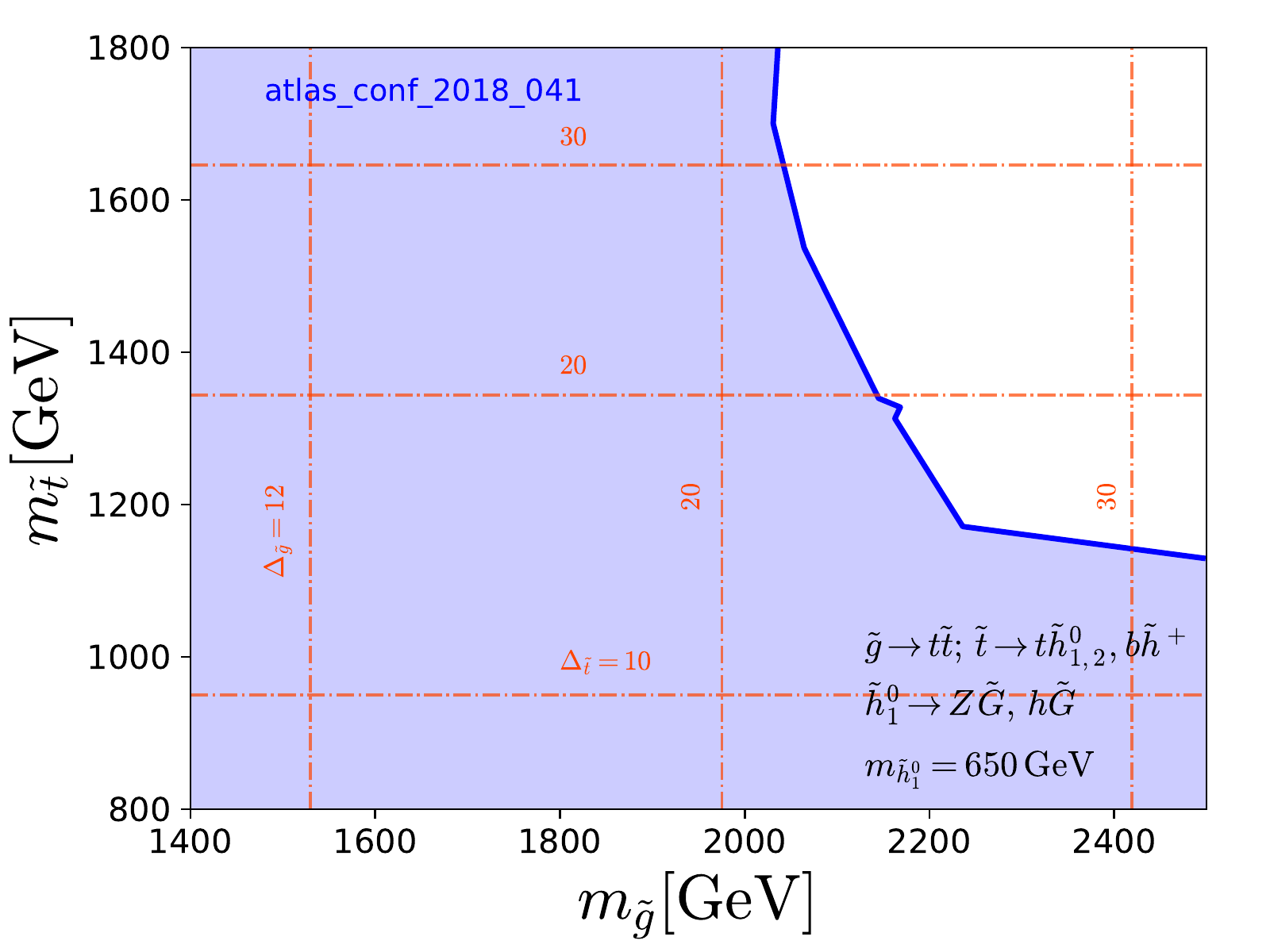}
        \end{center}
    \end{minipage}
    \caption{
    \small The exclusion plots in the stop-gluino model, Sec.~\ref{sec:stop-gluino}. The left column with electroweakino as the LSP and the right column with gravitino LSP. From top to bottom: bino, wino and higgsino case. Only the most constraining analyses are shown. For each plot the mass of the (N)LSP was assumed: $m_{\widetilde{B}} = 400$~GeV, $m_{\widetilde{W}} = 1$~TeV and $m_{\widetilde{h}} = 650$~GeV, respectively. In the higgsino scenario (bottom) lines of constant fine tuning were imposed: horizontal for top contribution and vertical for gluino contribution; see text for details.
    }
    \label{fig:gluino-stop}
\end{figure}

Finally we analyse a model that admits both stops and gluinos. Depending on the mass hierarchy there will be two possible decay chains for gluinos. If $m_{\gluino} > m_{\tone}+m_t$ the gluino will decay as $\gluino \to t \tone $. Otherwise the gluino will have a three-body decay with the pattern described in the previous section. The limits are summarized in Fig.~\ref{fig:gluino-stop}.  

In the left column of Fig.~\ref{fig:gluino-stop} we show the limits on the stop-gluino simplified model with electroweakino LSP in the  $(\gluino, \tone)$ plane for fixed LSP masses: $m_{\widetilde{B}} = 400$~GeV, $m_{\widetilde{W}} = 1$\,TeV and $m_{\widetilde{h}} = 650$\,GeV. While there is no experimental limit on the bino \cite{Dreiner:2009ic}, the values for winos and higgsinos are motivated by the direct searches, in particular  \verb|cms_1801_03957|. It turns out that regardless of the nature of the LSP the most constraining analysis is \verb|atlas_conf_2018_041| and the exclusion is very similar for all types of the electroweakino LSP. The gluinos of mass up to 1800--2100~GeV are excluded for any stop mass, while we do not observe additional exclusion for stops, which is consistent with findings from previous sections.

We now turn to stop-gluino model with the gravitino LSP. For the bino NLSP, as can be seen in top-right panel of Fig.~\ref{fig:gluino-stop}, the strongest limit again comes from  \verb|atlas_1802_03158| (GMSB;~ $\gamma( \ge 1) + E_T^{\rm miss}$) \cite{1802_03158} except for the small area where \verb|atlas_conf_2018_041| is slightly stronger. The limits consistent with the previous two-dimensional scenarios are obtained when one of the particles is very heavy; $m_{\gluino} \gtrsim 2000$ GeV for heavy $\tone$, and  $m_{\tone} \gtrsim 1300$ GeV for heavy $\gluino$. The limit is only slightly stronger if both of these particles are light. For example,  \verb|atlas_1802_03158| \cite{1802_03158} excludes $(m_{\gluino}, m_{\tone}) \sim (2000, 1600)$ GeV as can be seen in the Fig.~\ref{fig:gluino-stop}.

For the wino NLSP (right-middle panel of Fig.~\ref{fig:gluino-stop}) we observe the exclusion of gluino lighter than 2000~GeV for any stop mass. This limit is mostly driven by the \verb|atlas_conf_2018_041| search. For very light stop mass, $\sim 1$\,TeV, \verb|atlas_1802_03158| provides the strongest limit and excludes the gluino mass up to 2200 GeV. The lower bound on the stop mass in this scenario is given by 1 TeV, which directly comes from our assumption; $m_{\tone} > m_{\widetilde W} = 1$ TeV. Finally, in the higgsino NLSP case  (right-bottom panel of Fig.~\ref{fig:gluino-stop}) the most stringent limit is again provided by \verb|atlas_conf_2018_041|. The limits $m_{\gluino} > 2000$~GeV and $m_{\tone} > 1300$~GeV generally hold. A somewhat stronger exclusion is obtained in the intermediate mass regime for $2000 <m_{\gluino} < 2200$~GeV and $1200 <m_{\tone} < 1500$~GeV

For the scenarios with the higgsino (N)LSP we also show a contribution to the Higgs mass fine-tuning, $\Delta_x \equiv \frac{\delta m_h^2}{\delta m_x^2} \frac{m_x^2}{ m_h^2 }$ with the orange dashed line, where the naive leading-log approximation expression Eq.~\eqref{eq:tuning}  and a large cut-off scale, $\Lambda = 10^{18}$ GeV, have been used to evaluate $\Delta_x$. The horizontal (vertical) lines show the fine-tuning contribution due to stops (gluinos).  Keeping in mind that one has a constant Higgsino contribution, $\Delta_{\widetilde h} = m^2_{\widetilde h} / m_h^2 \simeq 27$, over the plane, the largest contribution comes from the gluino, which amounts $\Delta_{\widetilde g} \gtrsim 700$ due to the gluino mass limit $m_{\widetilde g} \gtrsim 1800$ GeV. The gluino contribution to the fine-tuning is particularly large for scenarios with high scale cut-off because of its square dependence, $\Delta_{\widetilde g} \propto \left( \log \frac{\Lambda}{m_h} \right)^2$. 

In the light gravitino scenario, the cut-off scale can be lowered, and we take $\Lambda = 100$ TeV as an example. Compared to the case without gravitino, the fine-tuning contribution from gluino and stop are significantly relaxed due to much smaller logarithmic factors. One can see that in the region $m_{\widetilde{t}_1} \lesssim 1600$ GeV and $m_{\widetilde g} \lesssim 2400$ GeV,  the fine-tuning contributions from stop and gluino can be less than 30, and a similar contribution will be due to higgsino $\Delta_{\widetilde h} \sim 27$ for $m_{\widetilde h} = 650$ GeV. We see in general that the LHC constraint on the coloured particle masses is stronger for light gravitino scenario. However, low fine-tuning still favours the light gravitino scenario due to the suppression on the logarithmic factors with smaller cut-off scale.

\section{Conclusions}

In this paper  we have compared the lower bounds on the gluino and stop masses in two  supersymmetric scenarios, one  with neutralino as the LSP and the other  one with neutralino decaying into gravitino.  The analysis of the latest LHC data is based on three simplified models, with bino, wino and higgsino as the (N)LSP and under the assumption that other than those four particles are irrelevant for the collider signatures.  For the gravitino LSP, the considered signatures are for prompt neutralino decays, so that obtained bounds apply  when gravitinos are very light, in the eV to keV mass range. Such light gravitinos are very interesting cosmologically and typical  for gauge mediation models with the messenger mass scale of order ${\cal O}(100-1000)$ TeV. One may expect that low messenger scale ameliorates the fine tuning problem of  the Higgs potential. 
 
Generically,  the lower limits on the stop and gluino masses obtained   with  the gravitino LSP are significantly stronger than for the neutralino LSP. This is due  to the fact that the  gravitino decay as the last step in the chain of decays gives  signatures which have  very low SM background.  One may therefore expect that our conclusions for the comparison of the two scenarios remain qualitatively valid beyond the simplified models studied in this paper.

The fine tuning in the Higgs potential has been discussed in the model with higgsino (N)LSP. In this case all three relevant mass parameters, higgsino, stop and gluino masses, can be fixed at their minimal allowed values following from our analysis. Although with gravitino LSP the lower limits  on  the  gluino and stop masses are stronger than  with higgsino LSP  for the same higgsino mass, in gauge mediation models the net effect on the fine tuning   in the Higgs potential is substantially ameliorated thanks to a very low messenger (cut-off) scale.


\section*{Acknowledgement}
The work of SP is partially supported by the Beethoven grant DEC-2016/23/G/ST2/04301.
The work of KS is partially supported by the National Science Centre, Poland, under research grants 2017/26/E/ST2/00135.
The work of KR is partially supported by the National Science Centre, Poland, under research grants 2015/19/D/ST2/03136.
This research was partially supported by the Munich Institute for Astro- and Particle Physics (MIAPP) of the DFG Excellence Cluster Origins (www.origins-cluster.de).



\end{document}